\documentclass[aps,pra,twocolumn,showpacs,nobibnotes,nofootinbib]{revtex4-2}

\usepackage[utf8]{inputenc}
\usepackage[T1]{fontenc}
\usepackage{lmodern}
\usepackage{physics}
\usepackage{xcolor}
\usepackage[colorlinks,citecolor=blue,linkcolor=magenta,hypertexnames=false]{hyperref}
\usepackage[nameinlink,capitalize]{cleveref}
\usepackage{graphicx}
\usepackage{epsfig}
\usepackage{color}
\usepackage{bm}
\usepackage{amsmath,amssymb,amsfonts,bbm,physics,mathtools}

\newcommand{\prlrunin}[1]{\textit{#1}---\ignorespaces}

\makeatletter
\let\original@bib@device\bib@device
\renewcommand{\bib@device}[2]{%
  \begingroup
    \let\addcontentsline\@gobblethree
    \original@bib@device{#1}{#2}%
  \endgroup
}
\makeatother

\begin{document}

\title{Relaxation-Rectified Anti-Jaynes--Cummings Cascades for Autonomous Fock-State Stabilization}
\author{Yan Liu}\email{neonorth@foxmail.com}
\author{Jiahua Li}\email{Contact author: huajia\_li@163.com}
\affiliation{School of Physics, Huazhong University of Science and
Technology, Wuhan 430074, People's Republic of China}

\date{\today}

\begin{abstract}
We propose an autonomous scheme for stabilizing prescribed Fock states in a Kerr cavity by rectifying anti-Jaynes--Cummings interactions with auxiliary-qubit relaxation. 
Full Lindblad simulations demonstrate steady states dominated by the Fock states $\ket{1}$ through $\ket{4}$, accompanied by pronounced Wigner negativity. 
An analytical birth--death model captures the steady-state populations and identifies the operating regime for stabilization. 
These results establish auxiliary-qubit relaxation as a resource for autonomous reservoir engineering and provide a route to nonclassical bosonic-state preparation in a single Kerr cavity and, through a boundary reservoir, in a Kerr-cavity chain.
\end{abstract}

\maketitle

\prlrunin{Introduction}
Nonvacuum Fock states of a bosonic mode are paradigmatic nonclassical states with a definite excitation number and characteristic non-Gaussian phase-space structures~\cite{PhysRevLett.87.050402,PhysRevLett.89.200402, PhysRevLett.96.213601,PRXQuantum.2.030204,Deleglise2008}.
Their preparation and stabilization provide important resources for fundamental studies of quantum mechanics, quantum metrology, and bosonic quantum information processing~\cite{Deng2024,PhysRevLett.134.180801,Cai2021FundRes,Ofek2016Nature,CampagneIbarcq2020,Putterman2025}.
Approaches to generating bosonic Fock states can be broadly classified into transient preparation~\cite{Varcoe2000,Brattke2001,Geremia2006,GonzalezTudela2015,Hofheinz2008,Hofheinz2009,PhysRevLett.125.093603,PhysRevResearch.2.033489,PhysRevLett.133.203605} and steady-state stabilization~\cite{Sayrin2011,Peaudecerf2013,Prado2014SteadyFock,PhysRevLett.115.180501,PhysRevA.93.060301,PhysRevLett.132.203602}. 
Transient protocols produce the target state at a selected time~\cite{Hofheinz2008,Hofheinz2009,PhysRevLett.125.093603,PhysRevResearch.2.033489,PhysRevLett.133.203605}, whereas stabilization protocols maintain it through active feedback~\cite{Sayrin2011} or engineered dissipation~\cite{PhysRevLett.115.180501,PhysRevA.93.060301,PhysRevLett.132.203602}. 
In particular, autonomous stabilization requires neither continuous measurement nor real-time feedback, but instead makes the target state an attractor of the open-system dynamics~\cite{Poyatos1996,Verstraete2009_DissipativeQE,Harrington2022,Kienzler2015,Shankar2013,LeghtasScience2015,Li2024}.

Bosonic Fock states have been prepared through coherent qubit--oscillator control~\cite{Meekhof1996,Brattke2001,Hofheinz2008,Hofheinz2009,Chu2018} and stabilized using measurement-based feedback~\cite{Geremia2006,Sayrin2011,Zhou2012,Peaudecerf2013}. 
Autonomous stabilization has been pursued using photon-number-selective interactions combined with engineered dissipation, including engineered atomic reservoirs, coupling to a
lossy auxiliary mode, a Josephson junction under direct-current voltage bias, and cascaded selective photon addition~\cite{Prado2014SteadyFock,PhysRevLett.115.180501,PhysRevA.93.060301,PhysRevLett.132.203602}.
Related studies have also engineered bounded anti-Jaynes--Cummings (AJC) interactions for steady Fock-state generation~\cite{Rossetti2014,Rosado2015} and shown that qubit relaxation can enhance photon production induced by parametrically activated counter-rotating processes~\cite{Zhukov2016,Remizov2017}. 
In the former case, the finite Fock-space structure is built into the engineered interaction, whereas the latter does not provide photon-number-selective termination at a prescribed Fock state.
However, these approaches do not use relaxation of the participating auxiliary qubits to rectify a Kerr-resolved cascade of AJC transitions into autonomous stabilization of a prescribed Fock state.

Here, we show that auxiliary-qubit relaxation provides precisely such a rectification mechanism.  
In a Kerr cavity, each auxiliary qubit resonantly addresses a photon-number-selective AJC transition, combining experimentally accessible Kerr nonlinearities and photon-number-resolved spectra~\cite{Kirchmair2013,Lu2023SNAIL,He2023,Andersson2025} with parametrically activated sideband couplings~\cite{Beaudoin2012,Strand2013,Lu2017ParametricCoupling}.
Subsequent qubit relaxation retains the added cavity photon while resetting the qubit, thereby converting the reversible exchange into a directed photon-raising process. 
Fast and unconditional reset of superconducting qubits through lossy resonators or parametrically activated interactions has been demonstrated in circuit quantum electrodynamics (QED) platforms~\cite{Reed2010APL,Magnard2018PRL,Zhou2021,Sunada2022PRApplied}.
Cascading these processes generates a finite photon-raising ladder that terminates at a prescribed Fock state.

Compared with previous autonomous Fock-state-stabilization schemes~\cite{Prado2014SteadyFock,PhysRevLett.115.180501,PhysRevA.93.060301,PhysRevLett.132.203602}, the present mechanism obtains photon-number selectivity from the Kerr spectrum and directionality from relaxation of the same qubits that mediate the raising transitions.
The Kerr-resolved photon-raising cascade also distinguishes our scheme from relaxation-enhanced counter-rotating photon generation by terminating the excitation flow at a prescribed photon number~\cite{Zhukov2016,Remizov2017}.
Full Lindblad simulations demonstrate steady states dominated by the Fock states $\ket{1}$ through $\ket{4}$ with pronounced Wigner negativity, while an analytical birth--death (BD) model identifies the parameter regime for stabilization.
The derivation of the reduced population dynamics, the effects of pure dephasing and auxiliary-qubit parameter inhomogeneity, and the experimental implementation and feasibility are presented in the Supplemental Material (SM)~\cite{supplementary2025}.
We further show in the End Matter that the same finite photon-raising ladder can serve as a boundary reservoir for a Kerr-cavity chain, thereby extending the autonomous Fock-state stabilization to indirectly coupled cavities.
This extension places the present mechanism in the broader setting of driven-dissipative nonlinear photonic lattices and dissipatively stabilized photonic many-body states~\cite{PhysRevA.97.013853,PhysRevLett.122.110405,Ma2019}.

\begin{figure}[htb]
\centerline{\includegraphics[width=1.0\linewidth]{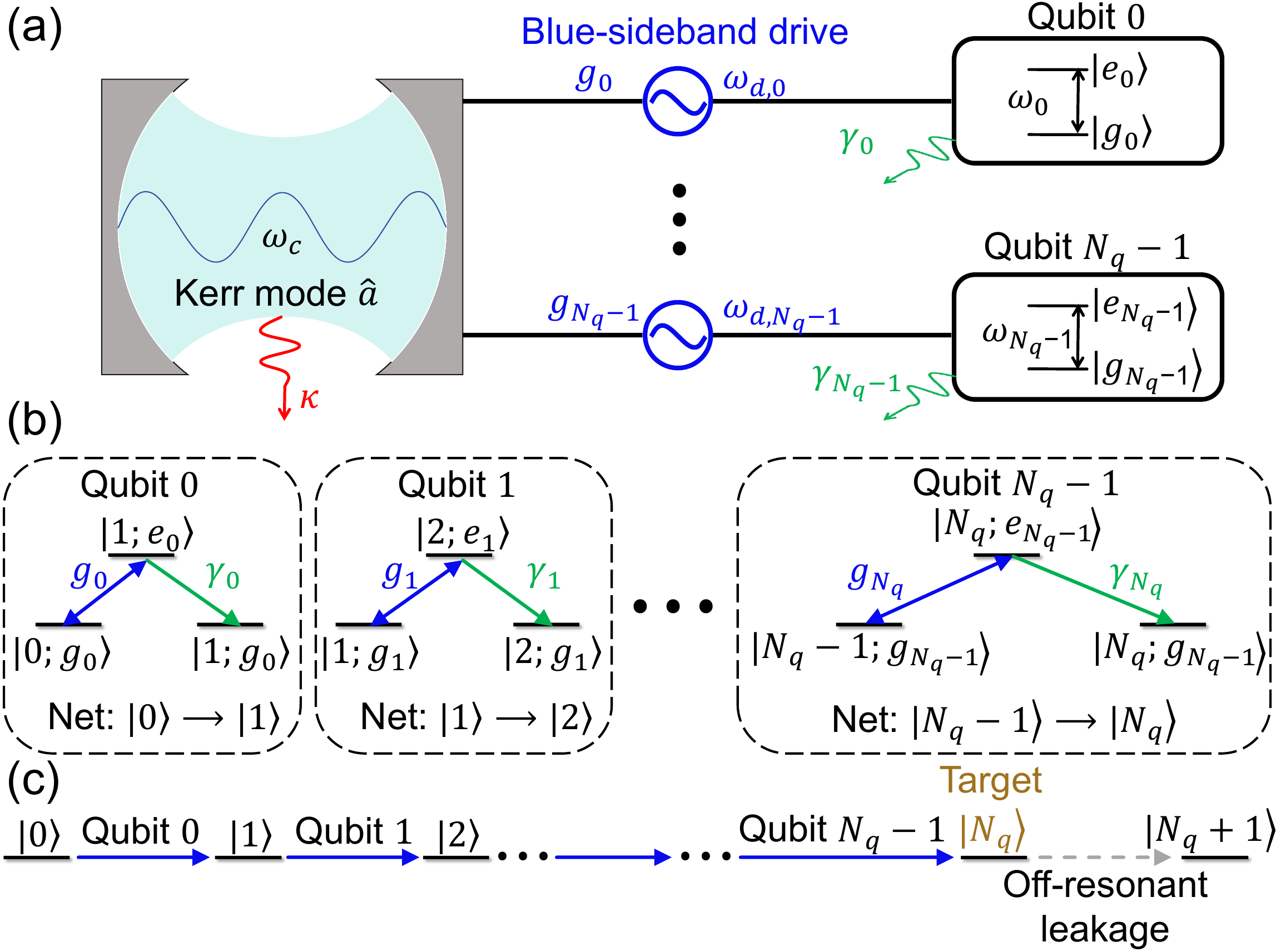}}
\caption{Dissipation-rectified Fock-state preparation.
(a) A Kerr cavity mode is coupled to $N_q$ relaxing auxiliary qubits through independently driven blue-sideband interactions. 
(b) For each qubit $\ell$, coherent coupling followed by qubit relaxation produces the net photon-raising transition $\lvert \ell\rangle\rightarrow\lvert \ell+1\rangle$.
(c) Cascading these processes pumps the cavity from $\lvert 0\rangle$ to the target state $\lvert N_q\rangle$, while further raising is off resonance.
Blue, green, and gray arrows denote coherent blue-sideband transitions or the resulting raising processes, qubit relaxation, and off-resonant leakage, respectively.
}
\label{fig1}
\end{figure}

\prlrunin{Model and dissipation-rectified photon raising}
As illustrated in Fig.\,\ref{fig1}(a), we consider a Kerr cavity mode coupled to an array of $N_q$ auxiliary qubits. 
The schematic in Fig.\,\ref{fig1} shows that the auxiliary qubits are labeled by $\ell=0,1,\ldots,N_q-1$.
The cavity mode has resonance frequency $\omega_c$, annihilation (creation) operator $\hat a$ ($\hat a^\dagger$), photon-number operator $\hat n=\hat a^\dagger\hat a$, Kerr nonlinearity $K$, and cavity loss rate $\kappa$.
The $\ell$-th qubit has transition frequency $\omega_\ell$, ground and excited states $|g_\ell\rangle$ and $|e_\ell\rangle$, and lowering (raising) operator $\hat\sigma_\ell^-=|g_\ell\rangle\langle e_\ell|$ ($\hat\sigma_\ell^+=|e_\ell\rangle\langle g_\ell|$).
A blue-sideband drive of frequency $\omega_{d,\ell}$ activates an effective counter-rotating interaction between the cavity and qubit $\ell$, with coupling strength $g_\ell$~\cite{Beaudoin2012,Strand2013}.

Moving to a frame rotating at $\omega_c$ for the cavity and at $\omega_{d,\ell}-\omega_c$ for qubit $\ell$, and setting $\hbar=1$, the effective Hamiltonian reads
\begin{equation}
\hat H
=
K\hat n^2
+
\sum_{\ell=0}^{N_q-1}
\Delta_\ell
\hat\sigma_\ell^+\hat\sigma_\ell^-
+
\sum_{\ell=0}^{N_q-1}
g_\ell
\left(
\hat a^\dagger\hat\sigma_\ell^+
+
\hat a\hat\sigma_\ell^-
\right),
\label{eq:H_AJC}
\end{equation}
where $\Delta_\ell=\omega_\ell+\omega_c-\omega_{d,\ell}$ is the effective blue-sideband detuning. 
A laboratory-frame derivation of Eq.\,\eqref{eq:H_AJC} from parametrically modulated transverse cavity--qubit couplings, together with the corresponding rotating-wave conditions and representative circuit-QED frequency scales, is given in Sec.\,S4 of the SM~\cite{supplementary2025}.
We denote the product basis of the cavity--qubit system by $|n;s_0,s_1,\ldots,s_{N_q-1}\rangle\equiv|n\rangle\bigotimes_{\ell=0}^{N_q-1}|s_\ell\rangle$, where $|n\rangle$ is the $n$-photon Fock state and $s_\ell\in\{g_\ell,e_\ell\}$.
For a process involving qubit $\ell$, we use the compact notation $|n,s_\ell;\mathbf{s}_{\bar{\ell}}\rangle$, where $\mathbf{s}_{\bar{\ell}}$ denotes the configuration of the spectator qubits. 
The spectator-qubit labels will be omitted when they are not essential.
The last term in Eq.\,(\ref{eq:H_AJC}) describes the engineered AJC interaction, which coherently couples $|n,g_\ell\rangle$ and $|n+1,e_\ell\rangle$.

Including cavity loss and qubit relaxation, the system density operator $\hat\rho$ obeys the Lindblad master equation
\begin{equation}
\frac{d\hat\rho}{dt}
=
-i[\hat H,\hat\rho]
+
\kappa\mathcal D[\hat a]\hat\rho
+
\sum_{\ell=0}^{N_q-1}
\gamma_\ell
\mathcal D[\hat\sigma_\ell^-]\hat\rho,
\label{eq:master}
\end{equation}
where $\gamma_\ell$ is the energy-relaxation rate of qubit $\ell$ and $\mathcal D[\hat O]\hat\rho=\hat O\hat\rho\hat O^\dagger-\frac{1}{2}(\hat O^\dagger\hat O\hat\rho+\hat\rho\hat O^\dagger\hat O)$ denotes the standard Lindblad dissipator~\cite{Gorini1976,Lindblad1976}.
For clarity, Eq.~\eqref{eq:master} retains only the dissipative channels essential to the rectification mechanism.
The effects of auxiliary-qubit and cavity pure dephasing are included and analyzed in Sec.\,S2 of the SM~\cite{supplementary2025}.

As illustrated in Fig.\,\ref{fig1}(b), qubit relaxation rectifies the coherent AJC exchange into a directed photon-raising process. 
Specifically, the local cycle $|n,g_\ell\rangle\xrightarrow{\,g_\ell}|n+1,e_\ell\rangle\xrightarrow{\,\gamma_\ell\,}|n+1,g_\ell\rangle$ resets the auxiliary qubit while retaining the additional photon in the cavity, yielding the net transition $|n\rangle\rightarrow|n+1\rangle$.
Thus, qubit relaxation converts the reversible counter-rotating exchange into an effectively irreversible photon-raising process.

The Kerr nonlinearity renders this process photon-number selective.
The energy mismatch for the AJC transition $|n,g_\ell\rangle\leftrightarrow|n+1,e_\ell\rangle$ is $\delta_{\ell,n}=\Delta_\ell+(2n+1)K$.
To resonantly address the transition $|\ell\rangle\rightarrow|\ell+1\rangle$, we choose $\Delta_\ell=-(2\ell+1)K+\delta_{\rm com}$, which gives $\delta_{\ell,n}=2(n-\ell)K+\delta_{\rm com}$, where $\delta_{\rm com}$ denotes a common detuning from the ideal resonance condition. 
For $\delta_{\rm com}=0$, the intended transition is resonant, whereas the remaining transitions driven by the same qubit are detuned by integer multiples of $2K$.

Consequently, as shown in Fig.\,\ref{fig1}(c), assigning $N_q$ auxiliary qubits, labeled by $\ell=0,1,\ldots,N_q-1$, to successive photon-raising transitions generates the finite photon-raising ladder $\ket{0}\rightarrow\ket{1}\rightarrow\cdots\rightarrow\ket{N_q}$.
Because no auxiliary qubit resonantly addresses the subsequent transition $\ket{N_q}\rightarrow\ket{N_q+1}$, the engineered photon-raising cascade terminates at $\ket{N_q}$, while further excitation above the target state remains off resonance.
The competition between the directed photon-raising processes and cavity photon loss then produces a steady state concentrated near the target Fock state $\ket{N_q}$.

To characterize the resulting cavity state, we trace out the auxiliary qubits and define the reduced cavity density operator $\hat\rho_c(t)=\mathrm{Tr}_{q}[\hat\rho(t)]$.
The population of the $n$-photon Fock state is $P_n(t)=\langle n|\hat\rho_c(t)|n\rangle$; its steady-state value is denoted by $P_n^{\rm ss}=\langle n|\hat\rho_c^{\rm ss}|n\rangle$.
The corresponding Wigner function is $W(\alpha)=\frac{2}{\pi}\mathrm{Tr}_c[\hat\rho_c\hat D(\alpha)\hat\Pi\hat D^\dagger(\alpha)]$, where $\hat D(\alpha)=\exp(\alpha\hat a^\dagger-\alpha^*\hat a)$ is the displacement operator and $\hat\Pi=(-1)^{\hat n}$ is the photon-number parity operator~\cite{Wigner1932,PhysRevA.15.449,Banaszek1996,
Lutterbach1997,PhysRevA.60.674}.
We quantify the total negative phase-space volume by the integrated Wigner negativity $\mathcal N_W=\frac{1}{2}\int d^2\alpha\,[|W(\alpha)|-W(\alpha)]$~\cite{Kenfack2004,qmjh-3n35,PhysRevA.98.052350,Chabaud2021WitnessingWN}.
Unless otherwise stated, we take $g_\ell=g$ and $\gamma_\ell=\gamma$ as a uniform reference setting and tune each qubit to resonance with its assigned AJC transition, $\Delta_\ell=-(2\ell+1)K$, corresponding to $\delta_{\rm com}=0$. 
Qubit-dependent parameter variations are examined in Sec.\,S3 of the SM~\cite{supplementary2025}.

\begin{figure}[!t]
\centering
\includegraphics[
    height=0.52\textheight,
    keepaspectratio
]{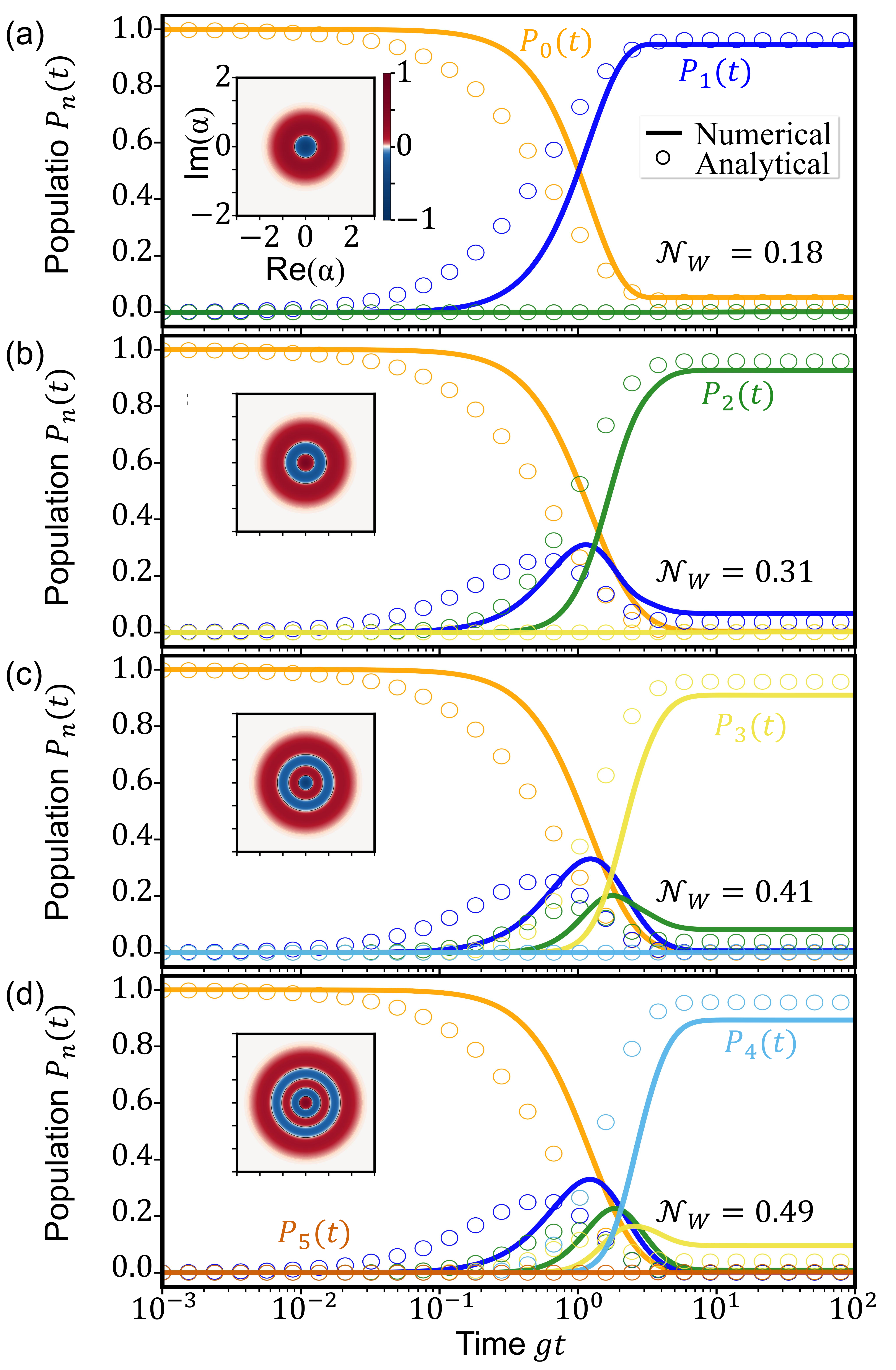}
\caption{Time evolution of the cavity Fock-state populations for target states $|1\rangle$, $|2\rangle$, $|3\rangle$, and $|4\rangle$ in panels (a)--(d), respectively.
Different colors denote the populations $P_n(t)$ of the corresponding Fock states $|n\rangle$.
Solid curves show the full numerical Lindblad dynamics, while open circles denote the corresponding analytical BD results.
Insets show the Wigner functions of the stabilized cavity states, with the corresponding Wigner negativities $\mathcal N_W$ indicated.
The parameters are $K/g=100$, $\kappa/g=0.05$, and $\gamma/g=3$.
}
\label{fig2}
\end{figure}

\prlrunin{Autonomous Fock-state stabilization}
We now show that the cascaded photon-raising processes autonomously drive the Kerr cavity into steady states dominated by prescribed Fock states.
Starting from the vacuum state with all auxiliary qubits in their ground states, we simulate the dynamics governed by Eq.\,\eqref{eq:master} for $N_q=1,\ldots,4$, with $\ket{N_q}$ as the target Fock state.
As shown by the solid curves in Fig.\,\ref{fig2}, the vacuum population is depleted and transferred successively along the Fock-state ladder before accumulating near the target state. 
For $N_q>1$, the intermediate populations exhibit transient maxima, directly revealing the sequential character of the cascaded photon-raising dynamics.

A notable feature of these dynamics is that the population is efficiently transferred toward the target Fock state even though all AJC couplings are chosen equal, $g_\ell=g$. 
This behavior originates in the bosonic creation operator $a^\dagger$: the matrix element for the transition $\ket{n,g_n}\leftrightarrow\ket{n+1,e_n}$ is $g\sqrt{n+1}$ and therefore increases with photon number. 
Consequently, the successive photon-raising steps do not become weaker as the population moves up the Fock-state ladder. 
Combined with relaxation-induced rectification, this bosonic enhancement enables efficient population transfer toward the target Fock state $\ket{N_q}$ without requiring step-dependent coupling strengths.

The population dynamics can be understood by adiabatically eliminating the auxiliary qubits, which reduces the full quantum dynamics to the nearest-neighbor BD equation
\begin{equation}
\dot P_n
=
\lambda_{n-1}P_{n-1}
+
\mu_{n+1}P_{n+1}
-
(\lambda_n+\mu_n)P_n ,
\label{eq:birth_death}
\end{equation}
where $\lambda_n$ and $\mu_n$ are the effective upward and downward transition rates, respectively.
Their derivation from the full Lindblad dynamics, together with the finite-dimensional time-dependent solution, the closed-form steady-state distribution, and a convergence check, is given in Sec.\,S1 of the SM~\cite{supplementary2025}.
In particular, the adjacent steady-state populations satisfy $P_{n+1}^{\rm ss}/P_n^{\rm ss}=\lambda_n/\mu_{n+1}$.
A distribution sharply concentrated near $\ket{N_q}$ is therefore obtained when this ratio is much larger than unity below the target and much smaller than unity at and above the target.
The analytical solutions (open circles in Fig.\,\ref{fig2}) capture the sequential population transfer along the Fock-state ladder, with transient deviations diminishing toward the steady state, consistent with the distinct dynamical and stationary conditions discussed in Sec.\,S1 of the SM~\cite{supplementary2025}.

\begin{figure}[htb]
\centerline{\includegraphics[width=1.0\linewidth]{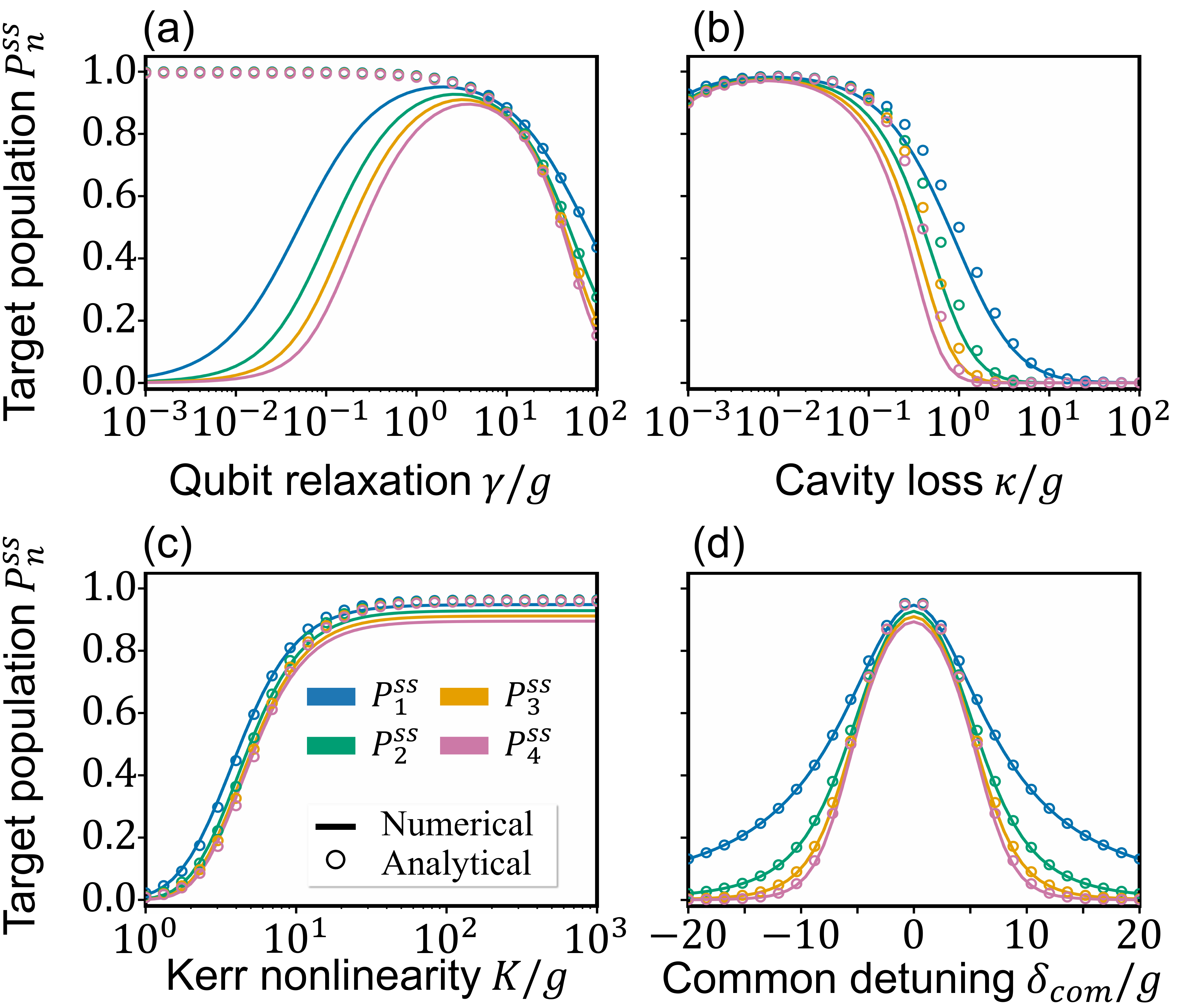}}
\caption{Steady-state target-state populations $P_n^{\rm ss}$ for $|n\rangle=|1\rangle,|2\rangle,|3\rangle$, and $|4\rangle$ as functions of (a) the qubit relaxation rate $\gamma/g$, (b) the cavity loss rate $\kappa/g$, (c) the Kerr nonlinearity $K/g$, and (d) the common detuning $\delta_{\rm com}/g$.
Different colors correspond to different target Fock states. 
Solid curves show the full numerical Lindblad results, while open circles denote the corresponding analytical BD results.
The fixed parameters are $K/g=100$, $\kappa/g=0.05$, and $\gamma/g=3$, except for the quantity varied in each panel.
In (c), the qubits are retuned according to $\Delta_\ell=-(2\ell+1)K$ as $K$ is varied; in (d), $\Delta_\ell=-(2\ell+1)K+\delta_{\rm com}$.
}
\label{fig3}
\end{figure}

At long times, the population becomes concentrated near the target Fock state $\ket{N_q}$ because the resonant photon-raising processes efficiently replenish population lost to lower Fock states, whereas excitation beyond $\ket{N_q}$ remains off resonance.
The modest reduction of the target-state population with increasing $N_q$ results from the photon-number-dependent loss rate $N_q\kappa$ together with accumulated off-resonant leakage. 
Nevertheless, the target state remains the dominant steady-state component for all cases considered, up to $\ket{4}$. 
The corresponding steady-state Wigner functions retain the characteristic concentric-ring structure and negative regions associated with the target Fock states, while the integrated Wigner negativity increases from $\mathcal{N}_W=0.18$ for $\ket{1}$ to $\mathcal{N}_W=0.49$ for $\ket{4}$. 
In principle, the target-state population could be further enhanced by introducing an additional unpumped auxiliary qubit, labeled $r$, with ground and excited states denoted by $\ket{g_{r}}$ and $\ket{e_{r}}$, respectively. 
By tuning its Jaynes--Cummings interaction into resonance with the transition $\ket{N_q+1,g_{r}}\leftrightarrow\ket{N_q,e_{r}}$, the resonant exchange followed by relaxation of this qubit at a suitable rate realizes an effective correction $\ket{N_q+1}\rightarrow\ket{N_q}$, thereby suppressing the leading leakage above the target state~\cite{Liu2026WignerNegative}.

\prlrunin{Parameter dependence of Fock-state stabilization}
We next examine how the steady-state target-state population depends on the principal coherent and dissipative parameters. 
Figure~\ref{fig3} shows the numerically obtained steady-state populations $P_n^{\rm ss}$ of the target Fock states $\ket{n}$ for $n=1,\ldots,4$, together with the corresponding predictions of the analytical BD model.
Except in the weak-qubit-relaxation regime, the close agreement between the solid curves and open circles over the parameter ranges considered confirms that the reduced model captures the dominant population-transfer processes governing the stabilization.

The nonmonotonic dependence on the qubit relaxation rate in Fig.\,\ref{fig3}(a) reflects the dual role of relaxation in rectifying and overdamping the coherent AJC transition.
For $\gamma\ll g$, the auxiliary qubits are reset too slowly following the coherent blue-sideband transitions, preventing the reversible cavity--qubit exchange from being effectively rectified into a unidirectional photon-raising process.
As $\gamma$ increases, faster qubit reset suppresses the reverse population flow and allows the photon-raising process to be repeated more efficiently, so that the target-state population initially increases and reaches a maximum at an intermediate relaxation rate.
For excessively large $\gamma$, however, the coherent cavity--qubit exchange becomes overdamped. 
As derived in the SM~\cite{supplementary2025}, in the resolved-transition regime the effective upward rate below the target is dominated by the resonant contribution $\lambda_n\simeq 2g^2(n+1)/\Gamma_n$, where $\Gamma_n$ is the total decay rate of the corresponding transition coherence. 
When qubit relaxation dominates the total decay rate, $\Gamma_n\simeq\gamma/2$, and hence
$\lambda_n\simeq4g^2(n+1)/\gamma$. 
At the same time, the increased total decay rate weakens the photon-number selectivity. Consequently, the target-state populations decrease again in the strong-relaxation regime.

The analytical BD model accurately captures the numerical results once the qubit reset is sufficiently fast, but substantially overestimates the target populations in the weak-relaxation regime. 
This discrepancy reflects the breakdown of the strong-relaxation assumption that the auxiliary qubits remain near their ground states.
For weak relaxation, this assumption is no longer valid: the qubits retain a non-negligible excited-state population, and coherent cavity--qubit exchange drives population back toward lower photon numbers.

The cavity-loss dependence in Fig.\,\ref{fig3}(b) is likewise governed by two competing effects. 
A finite cavity loss rate helps remove the small population transferred above the target state through off-resonant processes, producing a broad optimum at weak but nonzero $\kappa$.
As $\kappa$ is increased further, the photon-number-dependent decay rate $n\kappa$ overwhelms the engineered upward transfer and rapidly suppresses the target-state population. 
Higher Fock states are more sensitive to this competition because they experience larger direct loss rates and require a longer sequence of successful photon-raising steps.

As shown in Fig.\,\ref{fig3}(c), the steady-state target population increases with the Kerr nonlinearity. 
For small $K/g$, neighboring photon-raising transitions are not sufficiently separated in frequency, allowing off-resonant excitation above the target state and the associated population
leakage. 
Increasing $K/g$ suppresses this leakage by spectrally resolving the neighboring transitions.
Quantitatively, the unwanted transitions have detunings that are integer multiples of $2K$, and the corresponding off-resonant upward rates therefore decrease approximately as $K^{-2}$ (see the SM~\cite{supplementary2025} for the explicit rate expressions).
Once the neighboring transitions are well resolved, further increasing $K$ produces little additional improvement, and the target-state populations approach plateaus determined primarily by cavity photon loss and the finite resonant photon-raising rates.

\begin{figure}[htb]
\centerline{\includegraphics[width=1.0\linewidth]{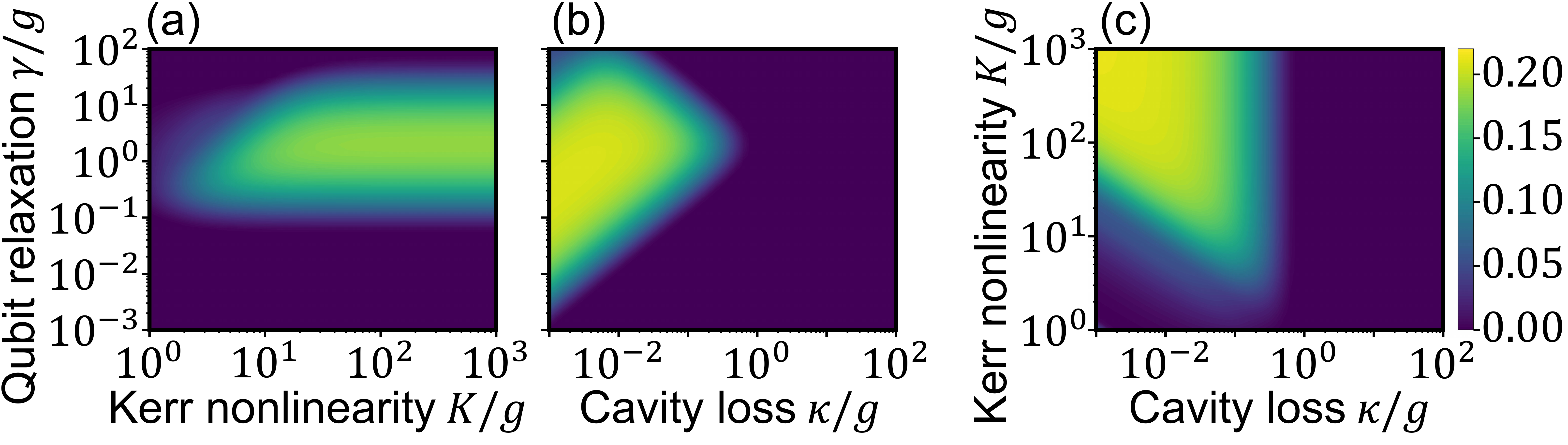}}
\caption{Steady-state Wigner negativity $\mathcal N_W$ of the stabilized single-photon state over different parameter planes.
(a) $\mathcal N_W$ versus $K/g$ and $\gamma/g$ at $\kappa/g=0.05$.
(b) $\mathcal N_W$ versus $\kappa/g$ and $\gamma/g$ at $K/g=100$.
(c) $\mathcal N_W$ versus $\kappa/g$ and $K/g$ at $\gamma/g=3$.
When $K$ is varied, the qubit is retuned according to $\Delta_0=-K$ to maintain resonance with the target $|0,g_0\rangle\leftrightarrow|1,e_0\rangle$ transition.
All panels share the same color scale.
}
\label{fig4}
\end{figure}

Finally, Fig.\,\ref{fig3}(d) examines the sensitivity of the steady-state target populations to a common detuning offset. 
A nonzero $\delta_{\rm com}$ places all engineered AJC transitions
off resonance, thereby reducing the corresponding effective upward rates.
Consequently, the target-state populations are maximal at $\delta_{\rm com}=0$ and decrease as $|\delta_{\rm com}|$ increases.
The approximately symmetric target-population profiles reflect the Lorentzian dependence of the dominant resonant upward rates on the common detuning (see the SM~\cite{supplementary2025}).
For larger $N_q$, the target-population peak becomes narrower because reaching $\ket{N_q}$ requires more successive AJC transitions, each of which is suppressed by the common detuning. 
The resulting reduction in the overall population-transfer efficiency becomes more pronounced as the number of required steps increases.

Except in the weak-qubit-relaxation regime, the analytical BD model closely reproduces the full Lindblad results, indicating that the steady-state target populations are controlled primarily by the competition among engineered resonant photon-raising processes, cavity-loss-induced downward transitions, and off-resonant excitation above the target state.

The same rate picture also explains the sensitivity to auxiliary-qubit and cavity pure dephasing analyzed in Sec.\,S2 of the SM~\cite{supplementary2025}.
Both dephasing channels increase the total decay rate $\Gamma_n$ of the number-selective transition coherences.
In the resolved-transition regime, this simultaneously suppresses the resonant upward transfer below the target and enhances residual off-resonant excitation above it, thereby reducing the target-state population.
The reduced BD model continues to closely reproduce the full Lindblad results.

\prlrunin{Steady-state Wigner negativity}
To complement the steady-state population analysis above, we characterize the stabilized single-photon state through its integrated Wigner negativity $\mathcal{N}_{W}$, which provides a direct phase-space witness of nonclassicality. 
As shown in Fig.\,\ref{fig4}, appreciable negativity occurs in the same parameter region in which the single-photon population is enhanced: it increases with the Kerr-induced photon-number selectivity, is maximized at an intermediate qubit relaxation rate, and is suppressed by cavity loss.
When varying $K$, the qubit is simultaneously retuned according to $\Delta_{0}=-K$, ensuring that the intended AJC transition $\ket{0,g_{0}}\leftrightarrow\ket{1,e_{0}}$ remains resonant. 
The dependence on $K$ therefore reflects improved spectral selectivity rather than a trivial
detuning of the target transition. 
In the optimal region, $\mathcal{N}_{W}$ approaches $0.20$, close to the value for an ideal single-photon Fock state, demonstrating that the parameters yielding a high steady-state single-photon population also produce strong phase-space nonclassicality.

\prlrunin{Conclusion}
We have shown that auxiliary-qubit relaxation rectifies parametrically activated AJC interactions into directed, number-selective photon-raising dynamics, enabling the autonomous stabilization of prescribed cavity Fock states.
The resulting steady states exhibit pronounced Wigner negativity, while the analytical BD description identifies the competition between resonant pumping, cavity loss, and off-resonant leakage that controls the stabilization.
The SM provides the corresponding rate derivation and stabilization criteria, convergence tests, robustness analyses, and implementation details~\cite{supplementary2025}.
Finally, the End Matter shows that the same rectification mechanism can serve as a boundary reservoir for a Kerr-cavity chain, extending autonomous Fock-state stabilization from the primary cavity to the indirectly coupled cavities and suggesting a route toward the autonomous preparation and repair of integer-filled bosonic lattice states.

\prlrunin{Acknowledgments}
We acknowledge Ying Wu and Xin-You L\"{u} for inspirational  discussions. 
The present research is supported in part by the National Natural Science Foundation of China (NSFC) through Grant No.\,12275092 and by the National Key Research and Development Program of China under Contract No.\,2021YFA1400700.

\prlrunin{Data availability}
The data supporting this study are available from the corresponding author upon reasonable request.

\bibliographystyle{apsrev4-2-bookfix-fixed}
\bibliography{WF-sub-arXiv}

\clearpage
\newpage
	\onecolumngrid
	\vspace{0.8cm}
	\begin{center}
		\rule{0.5\textwidth}{0.4pt}\\[6pt]
		{\large\textbf{End Matter}}\\[6pt]
		\rule{0.5\textwidth}{0.4pt}
	\end{center}
	\twocolumngrid

\begin{figure}[htb]
\centerline{\includegraphics[width=1.0\linewidth]{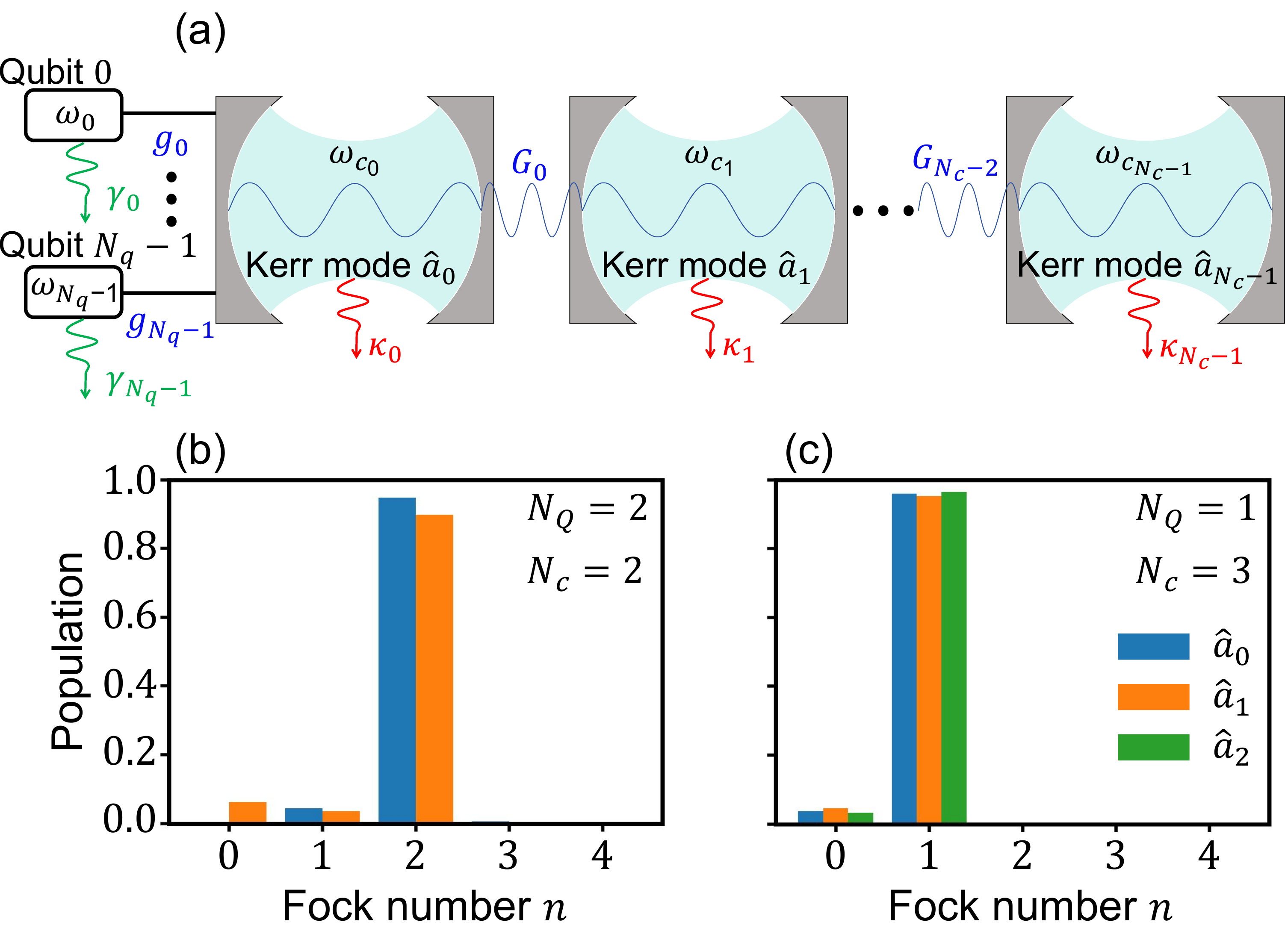}}
\caption{Extension of autonomous Fock-state stabilization to a Kerr-cavity chain.
(a) The primary mode $\hat{a}_0\equiv \hat{a}$, directly coupled to $N_q$ auxiliary qubits, is connected to $N_c-1$ identical lossy Kerr modes through nearest-neighbor beam-splitter interactions of strength $G$.
(b),(c) Local steady-state Fock-state populations of the individual cavities for $(N_q, N_c)=(2, 2)$ and $(1, 3)$, respectively.
In the indirectly coupled cavities, the population of the same target Fock state as in the primary cavity $\hat a_0$ is dominant, demonstrating the extension of the stabilization along the chain.
The parameters are $K/g=100$, $G/g=1$, $\gamma/g=3$, $\kappa/g=0.01$, and $\delta_{\mathrm{com}}/g=0$.
}
\label{fig5}
\end{figure}

\prlrunin{Extension to a Kerr-cavity chain}
The finite photon-raising ladder established in the main text can also function as a boundary reservoir for a Kerr-cavity chain.
As shown in Fig.\,\ref{fig5}(a), we extend the single-cavity setup to a chain of $N_c$ Kerr cavities labeled by $m=0,1,\ldots,N_c-1$. The annihilation and creation operators of cavity $m$ are denoted by $\hat a_m$ and $\hat a_m^\dagger$, respectively, with $\hat n_m=\hat a_m^\dagger\hat a_m$. 
The cavity of the single-cavity setup is relabeled as $\hat a_0\equiv\hat a$ and is referred to below as the primary cavity, whereas the auxiliary qubits remain labeled by $\ell=0,1,\ldots,N_q-1$.
Cavity $m$ has resonance frequency $\omega_{c,m}$, Kerr nonlinearity $K_m$, and cavity loss rate $\kappa_m$. 
Neighboring cavities $m$ and $m+1$ are coupled through a beam-splitter interaction of strength $G_m$. 
Only the primary cavity $\hat a_0$ is directly coupled to the auxiliary qubits, with AJC coupling strength $g_\ell$.
We consider identical cavity frequencies, $\omega_{c,m}=\omega_c$, and work in a frame rotating at $\omega_c$.
The Hamiltonian of the cavity--qubit chain is
\begin{align}
\hat H_{\mathrm{ch}}
={}&
\sum_{m=0}^{N_c-1}
K_m\hat n_m^{\,2}
+
\sum_{\ell=0}^{N_q-1}
\Delta_\ell
\hat\sigma_\ell^+\hat\sigma_\ell^-
\nonumber\\&
+
\sum_{\ell=0}^{N_q-1}
g_\ell
\left(
\hat a_0^\dagger\hat\sigma_\ell^+
+
\hat a_0\hat\sigma_\ell^-
\right)
\nonumber\\
&+
\sum_{m=0}^{N_c-2}
G_m
\left(
\hat a_m^\dagger\hat a_{m+1}
+
\hat a_{m+1}^\dagger\hat a_m
\right).
\label{eq:chain_H}
\end{align}
Including cavity loss and auxiliary-qubit relaxation, the density operator of the complete system obeys
\begin{equation}
\frac{d\hat\rho}{dt}
=
-i[\hat H_{\mathrm{ch}},\hat\rho]
+
\sum_{m=0}^{N_c-1}
\kappa_m\mathcal{D}[\hat a_m]\hat\rho
+
\sum_{\ell=0}^{N_q-1}
\gamma_\ell\mathcal{D}[\hat\sigma_\ell^-]\hat\rho .
\label{eq:chain_ME}
\end{equation}
For the results below, we take identical Kerr nonlinearities and cavity loss rates, $K_m=K$ and $\kappa_m=\kappa$, together with uniform nearest-neighbor couplings $G_m=G$. 
We also set $g_\ell=g$ and $\gamma_\ell=\gamma$ for all auxiliary qubits, while retaining the qubit-dependent detunings $\Delta_\ell=-(2\ell+1)K+\delta_{\mathrm{com}}$ required to address the successive transitions $\ket{\ell,g_\ell}\leftrightarrow\ket{\ell+1,e_\ell}$ of the primary cavity.

Figures~\ref{fig5}(b) and \ref{fig5}(c) show the local steady-state Fock-state populations of the individual cavities.
For $N_q=2$ and $N_c=2$, the two-photon Fock-state population is dominant in both the primary cavity $\hat a_0$ and the indirectly coupled cavity $\hat a_1$.
For $N_q=1$ and $N_c=3$, the single-photon Fock-state population remains dominant throughout the chain, including the terminal cavity $\hat a_2$.
Thus, although the auxiliary qubits act only on $\hat a_0$, nearest-neighbor photon exchange extends the autonomous Fock-state stabilization to the indirectly coupled Kerr cavities.

To clarify the underlying mechanism, we denote the joint cavity Fock state by $\ket{n_0,n_1,\ldots,n_{N_c-1}}\equiv\bigotimes_{m=0}^{N_c-1}\ket{n_m}_m$, where $n_m$ is the photon number in cavity $m$.
The auxiliary qubits provide a recurrent photon source at the primary cavity $\hat a_0$ located at the boundary of the chain: each AJC transition increases its photon number by one, while qubit relaxation resets the corresponding auxiliary qubit and allows the process to repeat. 
Because no auxiliary qubit resonantly addresses the transition from $\ket{N_q}_0$ to $\ket{N_q+1}_0$, the photon-raising cascade in the primary cavity terminates at the target Fock state $\ket{N_q}_0$.

For neighboring cavities with identical Kerr spectra, the beam-splitter interaction resonantly couples $\ket{n+1,n}_{m,m+1}$ and $\ket{n,n+1}_{m,m+1}$, with matrix element $G(n+1)$. 
This exchange alone is reversible and does not increase the total photon number. 
Here, however, the auxiliary qubits continuously replenish photons in the primary cavity $\hat a_0$. 
After one photon is transferred from $\hat a_m$ to $\hat a_{m+1}$ through the term $\hat a_{m+1}^\dagger\hat a_m$, the photon-raising process acting on the primary cavity $\hat a_0$ restores the photon removed from the boundary of the chain.
Repeated photon injection and nearest-neighbor exchange therefore act as an effective stepwise photon-raising process for the indirectly coupled cavities, allowing their target-state populations to build up successively along the chain.
This process terminates near the target photon number $N_q$. 
Once two neighboring cavities both reach the target photon number $N_q$, the beam-splitter interaction couples $\ket{N_q,N_q}_{m,m+1}$ to $\ket{N_q-1,N_q+1}_{m,m+1}$ and $\ket{N_q+1,N_q-1}_{m,m+1}$.
These transitions are detuned by $2K$ and are therefore strongly suppressed.
Moreover, because no auxiliary qubit resonantly drives $\ket{N_q}_0\rightarrow\ket{N_q+1}_0$, the primary cavity cannot supply the additional photon required to continue the resonant filling process. 
The Kerr nonlinearity and the termination of the photon-raising cascade therefore suppress populations above $\ket{N_q}$ throughout the cavity chain.

A single-photon loss from an indirectly coupled cavity produces a local photon-number deficit. 
The resonant exchange $\ket{N_q,N_q-1}_{m,m+1}\leftrightarrow\ket{N_q-1,N_q}_{m,m+1}$ coherently interconverts configurations in which the single-photon deficit resides in either of the two neighboring cavities. 
Repeated nearest-neighbor exchanges therefore allow a deficit created within the chain to reach the primary cavity, where the missing photon is restored by the qubit-induced raising process.
The combination of boundary photon injection, resonant nearest-neighbor exchange, and Kerr suppression above $N_q$ therefore accounts for the concentration of the steady-state Fock-state populations near $\ket{N_q}$ throughout the cavity chain.

From another perspective, the cavity chain realizes a recursive extension of the finite photon-raising ladder imposed by the auxiliary qubits. 
For the primary cavity, the $N_q$ qubits resonantly address the transitions $\ket{\ell}_0\rightarrow\ket{\ell+1}_0$ for $\ell=0,\ldots,N_q-1$, while qubit relaxation rectifies each transition into a directed photon-raising process.
For two neighboring cavities with identical Kerr spectra, the beam-splitter interaction resonantly couples $\ket{n+1,n}_{m,m+1}$ and $\ket{n,n+1}_{m,m+1}$, thereby transferring the same sequence of single-photon transitions from cavity $m$ to cavity $m+1$.
Crucially, the auxiliary qubits constitute the only source of photon-number increase in the cavity chain. 
Their resonant photon-raising channels terminate at $\ket{N_q}_0$, so no resonant process supplies an additional photon once the primary cavity reaches the target photon number. 
Since the beam-splitter interactions only redistribute photons between neighboring cavities, the finite photon-raising ladder inherited by each subsequent cavity likewise terminates effectively at $N_q$.
Consequently, cavity $m$ plays, for cavity $m+1$, an analogous functional role as the original set of $N_q$ auxiliary qubits for the primary cavity: it provides successive resonant single-photon transitions up to $\ket{N_q}_{m+1}$, but no resonant channel beyond it.
Repeating this structure recursively along the chain accounts for the extension of Fock-state stabilization to the indirectly coupled cavities.

The cavity chain can be implemented in a superconducting photonic lattice composed of Josephson-nonlinear microwave resonators with tunable nearest-neighbor beam-splitter couplings, while the auxiliary qubits are connected only to the primary cavity at the boundary~\cite{Houck2012,Gangat2013,Underwood2012,Fitzpatrick2017}.
Apart from a linear photon-number term, the onsite Kerr nonlinearity and nearest-neighbor photon hopping constitute a Bose--Hubbard-type lattice subject to local photon loss and boundary reservoir engineering~\cite{Hartmann2006,Greentree2006,Angelakis2007,
Carusotto2013,Noh2017}.
Related reservoir-engineering techniques have been used to stabilize a photonic Mott insulator in a superconducting Bose--Hubbard lattice~\cite{Ma2019}. 
The steady states obtained here, in which the population of the Fock state $\ket{N_q}$ dominates in each cavity, may therefore be viewed, at the level of local photon-number statistics, as finite-size nonequilibrium Mott-like photonic states.
This connection suggests a route to the autonomous preparation and repair of integer-filled bosonic lattice states using engineered dissipation applied only at the boundary.
Thus, the extension to a Kerr-cavity chain retains the relaxation-induced directionality and the Kerr-resolved termination of the photon-raising cascade.
Nearest-neighbor photon exchange then extends autonomous Fock-state stabilization from the primary cavity to the indirectly coupled cavities along the chain.

\nocite{
PhysRevA.85.032111,
Leek2009,
Tancredi2013,
FrattiniAPL2017,
Weissl2015,
PhysRevLett.110.120501,
Lu2021Decoherence,
Brehm2021,
Bolgar2020,
Cai2026KerrSqueezing,
Hoffman2011,
Roth2017}

\providecommand{\D}{\mathcal D}
\providecommand{\Tr}{\operatorname{Tr}}
\providecommand{\ImPart}{\operatorname{Im}}
\providecommand{\Hc}{\mathrm{H.c.}}
\providecommand{\ii}{\mathrm{i}}

%
\clearpage
\onecolumngrid

\setcounter{page}{1}
\setcounter{section}{0}
\setcounter{subsection}{0}
\setcounter{subsubsection}{0}
\setcounter{equation}{0}
\setcounter{figure}{0}
\setcounter{table}{0}

\renewcommand{\thesection}{S\arabic{section}}
\renewcommand{\theequation}{S\arabic{equation}}
\renewcommand{\thefigure}{S\arabic{figure}}
\renewcommand{\thetable}{S\arabic{table}}

\providecommand{\theHsection}{}
\providecommand{\theHsubsection}{}
\providecommand{\theHsubsubsection}{}
\providecommand{\theHequation}{}
\providecommand{\theHfigure}{}
\providecommand{\theHtable}{}
\renewcommand{\theHsection}{supp.section.\arabic{section}}
\renewcommand{\theHsubsection}{supp.subsection.\arabic{section}.\arabic{subsection}}
\renewcommand{\theHsubsubsection}{supp.subsubsection.\arabic{section}.\arabic{subsection}.\arabic{subsubsection}}
\renewcommand{\theHequation}{supp.equation.\arabic{equation}}
\renewcommand{\theHfigure}{supp.figure.\arabic{figure}}
\renewcommand{\theHtable}{supp.table.\arabic{table}}

\begingroup
\renewcommand{\thefootnote}{\fnsymbol{footnote}}
\begin{center}
{\Large\bfseries Supplemental Material for\\[0.35em]
``Relaxation-Rectified Anti-Jaynes--Cummings Cascades for Autonomous Fock-State Stabilization''\par}
\end{center}
\endgroup

\vspace{1.0em}


For convenience, the principal notation used throughout the Supplemental Material is summarized below. 
Quantities introduced only locally are defined at their first occurrence.
\begingroup
\small
\begin{center}
\begin{tabular}{@{}ll@{}}
\hline\hline
Symbol & Definition \\
\hline
$\hat a$, $\hat n=\hat a^\dagger\hat a$
& Cavity annihilation and photon-number operators \\

$\ell$, $n$
& Auxiliary-qubit index and cavity photon number \\

$N_q$
& Number of auxiliary qubits and target photon number \\

$N_{\mathrm{max}}$
& Upper photon-number cutoff of the reduced rate model \\

$K$
& Cavity Kerr coefficient in the convention $K\hat n^2$ \\

$g_\ell$
& Anti-Jaynes--Cummings coupling coefficient for qubit $\ell$ \\

$\Delta_\ell$
& Effective blue-sideband detuning of qubit $\ell$ \\

$\kappa$, $\gamma_\ell$
& Cavity-loss and qubit-relaxation rates \\

$\kappa_\phi$, $\gamma_{\phi,\ell}$
& Cavity and qubit pure-dephasing rates \\

$\delta_{\ell,n}$
& Detuning of the transition
$\ket{n,g_\ell}\leftrightarrow\ket{n+1,e_\ell}$ \\

$\Gamma_{\ell,n}$
& Decay rate of the corresponding transition coherence \\

$R_{\ell,n}$
& Effective exchange coefficient of that transition \\

$P_n$
& Reduced cavity population of the Fock state $\ket n$ \\

$\lambda_n$, $\mu_n$
& Effective upward and downward population rates \\

$r_n=P_{n+1}^{\mathrm{ss}}/P_n^{\mathrm{ss}}$
& Ratio of adjacent steady-state populations \\
\hline\hline
\end{tabular}
\end{center}
\endgroup

\section{Analytical birth--death model for cavity Fock-state populations}
\label{sec:analytical_rate_model}

In this section, we derive an effective rate equation for the cavity Fock-state populations from the full driven-dissipative dynamics introduced in the main text. 
We first eliminate the coherence associated with each number-selective anti-Jaynes--Cummings (AJC) transition and obtain the corresponding effective exchange coefficient. 
We then eliminate the auxiliary-qubit degrees of freedom in the strong-relaxation regime and reduce the cavity dynamics to a finite birth--death (BD) equation. 
The resulting model provides both the time-dependent populations and their closed-form steady-state distribution.

\subsection{Dissipative model and number-selective AJC transitions}
\label{sec:sm_model}

We adopt the notation of the main text and set $\hbar=1$. 
In the driven rotating frame, the Hamiltonian is
\begin{equation}
\hat H
=
K\hat n^2
+
\sum_{\ell=0}^{N_q-1}
\Delta_\ell
\hat\sigma_\ell^+\hat\sigma_\ell^-
+
\sum_{\ell=0}^{N_q-1}
g_\ell
\left(
\hat a^\dagger\hat\sigma_\ell^+
+
\hat a\hat\sigma_\ell^-
\right),
\label{eq:sm_H}
\end{equation}
where all quantities are defined in the main text.
For completeness, we consider a more general dissipative environment by including pure dephasing of both the cavity and the auxiliary qubits.
The corresponding master equation reads
\begin{align}
\frac{d\hat\rho}{dt}
=
-i[\hat H,\hat\rho]
+
\kappa\mathcal D[\hat a]\hat\rho
+
\sum_{\ell=0}^{N_q-1}
\gamma_\ell
\mathcal D[\hat\sigma_\ell^-]\hat\rho
+
\kappa_\phi\mathcal D[\hat n]\hat\rho
+
\sum_{\ell=0}^{N_q-1}
\gamma_{\phi,\ell}
\mathcal D[\hat\sigma_\ell^z]\hat\rho,
\label{eq:sm_master}
\end{align}
where $\hat\sigma_\ell^z=|e_\ell\rangle\langle e_\ell|-|g_\ell\rangle\langle g_\ell|$, and $\kappa_\phi$ and $\gamma_{\phi,\ell}$ denote the cavity and qubit pure-dephasing rates, respectively. With the convention used in
Eq.\,\eqref{eq:sm_master}, the term $\gamma_{\phi,\ell}\mathcal D[\hat\sigma_\ell^z]$ causes the qubit coherence to decay at the rate $2\gamma_{\phi,\ell}$.

The energy mismatch associated with the AJC transition $|n,g_\ell\rangle\leftrightarrow|n+1,e_\ell\rangle$ is
\begin{equation}
\delta_{\ell,n}
=
\Delta_\ell+(2n+1)K.
\label{eq:sm_delta}
\end{equation}
For the detuning choice $\Delta_\ell=-(2\ell+1)K+\delta_{\rm com}$ used in the main text,
Eq.\,\eqref{eq:sm_delta} becomes
\begin{equation}
\delta_{\ell,n}
=
2(n-\ell)K+\delta_{\rm com}.
\label{eq:sm_delta_selected}
\end{equation}
Thus, for $\delta_{\rm com}=0$, qubit $\ell$ is resonant with the selected transition $|\ell,g_\ell\rangle\leftrightarrow|\ell+1,e_\ell\rangle$, whereas the transition $|n,g_\ell\rangle\leftrightarrow|n+1,e_\ell\rangle$ with $n\neq\ell$ is detuned by $2(n-\ell)K$.

\subsection{Elimination of the transition coherence}
\label{sec:sm_coherence}

To derive the effective exchange coefficient associated with the transition $\ket{n,g_\ell}\leftrightarrow\ket{n+1,e_\ell}$ mediated by qubit $\ell$, we consider the reduced density operator of the cavity and qubit $\ell$,
\begin{equation}
\hat\rho_{c\ell}
=
\Tr_{\bar\ell}\hat\rho,
\label{eq:sm_reduced_state}
\end{equation}
where $\Tr_{\bar\ell}$ denotes the trace over all auxiliary qubits except qubit $\ell$. 
Using the local-state notation introduced in the main text, we denote the two states of this reduced subsystem by
\begin{equation}
\ket{A}
\equiv
\ket{n,g_\ell},
\qquad
\ket{B}
\equiv
\ket{n+1,e_\ell}.
\end{equation}
The corresponding reduced populations and transition coherence are
\begin{equation}
P_A=\mel{A}{\hat\rho_{c\ell}}{A},
\qquad
P_B=\mel{B}{\hat\rho_{c\ell}}{B},
\qquad
C_{\ell,n}=\mel{A}{\hat\rho_{c\ell}}{B}.
\label{eq:sm_local_variables}
\end{equation}

The part of the cavity--qubit-$\ell$ Hamiltonian $\hat H$ in Eq.\,\eqref{eq:sm_H} that couples the two local states $\ket{A}$ and $\ket{B}$ is
\begin{equation}
\hat H_{\ell,n}
=
E_{A,n}|A\rangle\langle A|
+
E_{B,\ell,n}|B\rangle\langle B|
+
g_\ell\sqrt{n+1}
\left(
|B\rangle\langle A|
+
|A\rangle\langle B|
\right),
\label{eq:sm_local_H}
\end{equation}
where
\begin{equation}
E_{A,n}=Kn^2,
\qquad
E_{B,\ell,n}=K(n+1)^2+\Delta_\ell,
\end{equation}
and hence
\begin{equation}
E_{B,\ell,n}-E_{A,n}
=
\Delta_\ell+(2n+1)K
=
\delta_{\ell,n}.
\label{eq:sm_local_energy_difference}
\end{equation}

Here and below, an overdot denotes a time derivative, $\dot{X}\equiv dX/dt$, while $\left.\dot{X}\right|_H$ denotes the contribution to the time derivative of $X$ arising from the Hamiltonian part of the master equation.
Accordingly, the contribution of $\hat H_{\ell,n}$ to the reduced density-operator dynamics is
\begin{equation}
\left.\dot{\hat\rho}_{c\ell}\right|_H
=
-i[\hat H_{\ell,n},\hat\rho_{c\ell}].
\label{eq:sm_rho_H}
\end{equation}
Using $P_A=\mel{A}{\hat\rho_{c\ell}}{A}$, $P_B=\mel{B}{\hat\rho_{c\ell}}{B}$, and $C_{\ell,n}=\mel{A}{\hat\rho_{c\ell}}{B}$, we obtain
\begin{align}
\left.\dot P_A\right|_H
&=
-i\langle A|[\hat H_{\ell,n},\hat\rho_{c\ell}]|A\rangle
=-2g_\ell\sqrt{n+1}\operatorname{Im}C_{\ell,n},
\label{eq:sm_PA_current}
\\
\left.\dot P_B\right|_H
&=
-i\langle B|[\hat H_{\ell,n},\hat\rho_{c\ell}]|B\rangle
=+2g_\ell\sqrt{n+1}\operatorname{Im}C_{\ell,n}.
\label{eq:sm_PB_current}
\end{align}

After tracing Eq.\,\eqref{eq:sm_master} over the remaining auxiliary qubits, the exact equation for $C_{\ell,n}$ generally contains additional terms involving correlations generated by the couplings to the other qubits.
For the detuning pattern considered here, the AJC channels associated with two different auxiliary qubits $\ell$ and $j$ are separated by
\begin{equation}
\left|\delta_{\ell,n}-\delta_{j,n}\right|
=
\left|\Delta_\ell-\Delta_j\right|
=
2|\ell-j||K|,
\qquad j\neq\ell .
\end{equation}
Thus, the minimum separation between distinct qubit-resolved channels is $2|K|$.
We work in the resolved-transition regime in which this Kerr-induced spectral separation is much larger than the relevant coherent-coupling and dissipative scales, namely,
\begin{equation}
2|K|
\gg
g_j\sqrt{n+1},\;
\gamma_j,\;
(2n+1)\kappa,\;
\gamma_{\phi,j},\;
\kappa_\phi ,
\end{equation}
for the auxiliary qubits and photon numbers relevant to the stabilization dynamics.
Under this hierarchy, cross-qubit coherences generated by distinct AJC channels are strongly off resonant and their feedback on $C_{\ell,n}$ is perturbatively suppressed.
We therefore neglect these cross-qubit correlation terms and close the equation at the cavity--qubit-$\ell$ level.
The retained terms consist of the cavity--qubit-$\ell$ Hamiltonian contribution and the local dissipative contributions. 
The resulting equation for $C_{\ell,n}$ can be decomposed as
\begin{equation}
\dot C_{\ell,n}
=
\left.\dot C_{\ell,n}\right|_{H}
+
\left.\dot C_{\ell,n}\right|_{\gamma_\ell}
+
\left.\dot C_{\ell,n}\right|_{\kappa}
+
\left.\dot C_{\ell,n}\right|_{\gamma_{\phi,\ell}}
+
\left.\dot C_{\ell,n}\right|_{\kappa_\phi},
\label{eq:sm_C_decomposition}
\end{equation}
where $\left.\dot C_{\ell,n}\right|_{X}$ denotes the contribution to
$\dot C_{\ell,n}$ generated by the term associated with $X$ in the master
equation.
The Hamiltonian contribution is
\begin{align}
\left.\dot C_{\ell,n}\right|_{H}
&=
-i\langle A|[\hat H_{\ell,n},\hat\rho_{c\ell}]|B\rangle
=
i\delta_{\ell,n}C_{\ell,n}
-i g_\ell\sqrt{n+1}\left(P_B-P_A\right).
\label{eq:sm_C_H}
\end{align}
The contribution from qubit relaxation is
\begin{align}
\left.\dot C_{\ell,n}\right|_{\gamma_\ell}
&=
\gamma_\ell
\langle A|
\mathcal D[\hat\sigma_\ell^-]\hat\rho_{c\ell}
|B\rangle
=
-\frac{\gamma_\ell}{2}C_{\ell,n}.
\label{eq:sm_C_gamma}
\end{align}
The cavity-loss contribution is
\begin{align}
\left.\dot C_{\ell,n}\right|_{\kappa}
=
\kappa
\langle A|
\mathcal D[\hat a]\hat\rho_{c\ell}
|B\rangle
=
\kappa\sqrt{(n+1)(n+2)}\,C_{\ell,n+1}
-\frac{(2n+1)\kappa}{2}C_{\ell,n},
\label{eq:sm_C_kappa}
\end{align}
where $C_{\ell,n+1}=\langle n+1,g_\ell|\hat\rho_{c\ell}|n+2,e_\ell\rangle$ is the coherence associated with the neighboring AJC transition.
The qubit pure-dephasing contribution is
\begin{align}
\left.\dot C_{\ell,n}\right|_{\gamma_{\phi,\ell}}
=
\gamma_{\phi,\ell}
\langle A|
\mathcal D[\hat\sigma_\ell^z]\hat\rho_{c\ell}
|B\rangle
=
-2\gamma_{\phi,\ell}C_{\ell,n},
\label{eq:sm_C_qphi}
\end{align}
while cavity pure dephasing gives
\begin{align}
\left.\dot C_{\ell,n}\right|_{\kappa_\phi}
=
\kappa_\phi
\langle A|
\mathcal D[\hat n]\hat\rho_{c\ell}
|B\rangle
=
-\frac{\kappa_\phi}{2}C_{\ell,n}.
\label{eq:sm_C_cphi}
\end{align}

In the parameter regime considered here, the large Kerr nonlinearity not only spectrally separates neighboring number-selective transitions, but also suppresses the corresponding off-resonant transition coherences.
For a resonant or leading near-resonant coherence $C_{\ell,n}$, the detuning $|\delta_{\ell,n}|$ is small, whereas the neighboring coherence satisfies $\delta_{\ell,n+1}=\delta_{\ell,n}+2K$.
Hence, in the large-Kerr regime, $C_{\ell,n+1}$ is associated with a transition detuned by approximately $2K$ from the retained resonant or near-resonant channel.
Its amplitude is consequently suppressed when this neighboring transition is well resolved, while $C_{\ell,n}$ provides the leading coherence contribution.
Moreover, the feedback of $C_{\ell,n+1}$ onto $C_{\ell,n}$ enters only through the cavity-loss-induced term $\kappa\sqrt{(n+1)(n+2)}\,C_{\ell,n+1}$.
A convenient sufficient hierarchy (but not a necessary one) for treating this feedback as a subleading correction is
\begin{equation}
\kappa\sqrt{(n+1)(n+2)}
\ll
2|K|,
\end{equation}
together with the requirement that the neighboring transition remain well off resonant on the relevant coherent and dissipative scales.
Under these conditions, the Kerr-induced detuning suppresses the neighboring coherence, while its cavity-loss-induced coupling back into $C_{\ell,n}$ remains weak compared with the spectral separation, so that the resulting feedback constitutes only a perturbative correction.
We therefore neglect the feeding term proportional to $C_{\ell,n+1}$ in Eq.\,\eqref{eq:sm_C_kappa}.
The accuracy of this approximation is assessed below by comparison with the full Lindblad dynamics.
Combining the remaining contributions gives
\begin{equation}
\dot C_{\ell,n}
=
-
\left(
\Gamma_{\ell,n}
-i\delta_{\ell,n}
\right)
C_{\ell,n}
-i g_\ell\sqrt{n+1}\left(P_B-P_A\right),
\label{eq:sm_coherence_eq}
\end{equation}
where
\begin{equation}
\Gamma_{\ell,n}
=
\frac{\gamma_\ell}{2}
+
\frac{(2n+1)\kappa}{2}
+
2\gamma_{\phi,\ell}
+
\frac{\kappa_\phi}{2}
\label{eq:sm_Gamma}
\end{equation}
is the total decay rate of the coherence $C_{\ell,n}$.

For the time-dependent dynamics, the coherence $C_{\ell,n}$ can be adiabatically eliminated when its characteristic evolution is much faster than the population dynamics.
A convenient sufficient condition is
\begin{equation}
g_\ell\sqrt{n+1}
\ll
\sqrt{\Gamma_{\ell,n}^{2}+\delta_{\ell,n}^{2}}.
\end{equation}
For the resonant and near-resonant transitions that dominate the stabilization dynamics, this condition reduces to
$g_\ell\sqrt{n+1}\ll\Gamma_{\ell,n}$.
Under this timescale-separation condition, we set
$\dot C_{\ell,n}\simeq0$~\cite{PhysRevA.85.032111}, which gives
\begin{equation}
C_{\ell,n}
\simeq
\frac{
-i g_\ell\sqrt{n+1}(P_B-P_A)
}{
\Gamma_{\ell,n}-i\delta_{\ell,n}
}.
\label{eq:sm_C_eliminated}
\end{equation}
For the moderate relaxation rate used in the numerical calculations, this adiabatic condition is not asymptotically satisfied for all appreciably populated transitions.
Consequently, Eq.\,\eqref{eq:sm_C_eliminated} provides an approximate closure for the transient dynamics, and quantitative deviations from the full Lindblad evolution can occur when the timescale separation is only moderate.
Indeed, as shown in Fig.\,2 of the main text, visible deviations between the analytical and full Lindblad results occur during parts of the transient evolution, whereas the agreement becomes considerably better as the system approaches the steady state.
In the steady state, however, $\dot C_{\ell,n}=0$ holds exactly for the reduced local-coherence equation~\eqref{eq:sm_coherence_eq}. 
This does not imply that the eliminated coherence is exactly given by Eq.\,\eqref{eq:sm_C_eliminated} in the full Lindblad dynamics.
Equation~\eqref{eq:sm_C_eliminated} therefore follows directly in the steady state without requiring the adiabatic condition, with all quantities evaluated at stationarity.
This explains why the reduced model can reproduce the steady-state populations more accurately than the transient dynamics even when the dynamical adiabatic condition is only moderately satisfied.
The imaginary part of $C_{\ell,n}$ is
\begin{equation}
\operatorname{Im}C_{\ell,n}
=
\frac{
g_\ell\sqrt{n+1}\Gamma_{\ell,n}
}{
\Gamma_{\ell,n}^2+\delta_{\ell,n}^2
}
(P_A-P_B).
\label{eq:sm_ImC}
\end{equation}
Substituting the imaginary part of $C_{\ell,n}$ into Eqs.\,\eqref{eq:sm_PA_current} and \eqref{eq:sm_PB_current}, we obtain
\begin{equation}
\left.\dot P_B\right|_H
=
-\left.\dot P_A\right|_H
=
\frac{
2g_\ell^2(n+1)\Gamma_{\ell,n}
}{
\Gamma_{\ell,n}^2+\delta_{\ell,n}^2
}
(P_A-P_B).
\label{eq:sm_population_transfer}
\end{equation}
We therefore define the population current from $|A\rangle$ to $|B\rangle$ as
\begin{equation}
J_{\ell,n}
\equiv
\left.\dot P_B\right|_H
=
-\left.\dot P_A\right|_H
=
R_{\ell,n}(P_A-P_B),
\label{eq:sm_local_current}
\end{equation}
where
\begin{equation}
R_{\ell,n}
=
\frac{
2g_\ell^2(n+1)\Gamma_{\ell,n}
}{
\Gamma_{\ell,n}^2+\delta_{\ell,n}^2
}
\label{eq:sm_R}
\end{equation}
is the effective exchange coefficient associated with the transition $|n,g_\ell\rangle\leftrightarrow|n+1,e_\ell\rangle$.

Equation~\eqref{eq:sm_R} also reveals the role of the total coherence decay rate $\Gamma_{\ell,n}$ in the number-selective transition.
For a resonant transition, $\delta_{\ell,n}=0$, and the effective exchange coefficient reduces to
\begin{equation}
R_{\ell,n}^{\rm res}
=
\frac{2g_\ell^2(n+1)}{\Gamma_{\ell,n}}.
\label{eq:sm_Rres}
\end{equation}
By contrast, for an off-resonant transition satisfying $|\delta_{\ell,n}|\gg\Gamma_{\ell,n}$, the effective exchange coefficient is
\begin{equation}
R_{\ell,n}^{\rm off}
\simeq
\frac{
2g_\ell^2(n+1)\Gamma_{\ell,n}
}{
\delta_{\ell,n}^2
}.
\label{eq:sm_Roff}
\end{equation}
Thus, increasing $\Gamma_{\ell,n}$ suppresses the resonant transition while enhancing off-resonant transfer, thereby reducing the photon-number selectivity.
Pure dephasing does not directly change the Fock-state populations, but modifies their dynamics through its contribution to $\Gamma_{\ell,n}$ and hence to the effective exchange coefficient $R_{\ell,n}$.

\subsection{Reduced birth--death equation}
\label{sec:sm_birth_death}

We next eliminate the auxiliary-qubit degrees of freedom from the population dynamics.
Let $Q_{g,\ell}$ and $Q_{e,\ell}$ denote the ground- and excited-state populations of qubit $\ell$, respectively, with
\begin{equation}
Q_{g,\ell}+Q_{e,\ell}=1.
\end{equation}
Neglecting population correlations between the cavity and qubit $\ell$, we use the factorized approximation
\begin{equation}
P_A\simeq Q_{g,\ell}P_n,
\qquad
P_B\simeq Q_{e,\ell}P_{n+1},
\label{eq:sm_population_closure}
\end{equation}
where $P_n(t)=\langle n|\hat\rho_c(t)|n\rangle$ is the cavity Fock-state population defined in the main text.
This approximation neglects cavity--auxiliary-qubit population correlations generated during the stabilization process.

Substituting Eq.\,\eqref{eq:sm_population_closure} into Eq.\,\eqref{eq:sm_local_current} gives
\begin{equation}
J_{\ell,n}
=
R_{\ell,n}
\left(
Q_{g,\ell}P_n
-
Q_{e,\ell}P_{n+1}
\right).
\label{eq:sm_reduced_current}
\end{equation}
The first and second terms describe the forward $\ket{n}\rightarrow \ket{n+1}$ and reverse $\ket{n+1}\rightarrow \ket{n}$ population-transfer processes, respectively.
Qubit relaxation biases these two processes by maintaining a larger ground-state population. 
In the strong-relaxation limit, where $\gamma_\ell\gg g_\ell\sqrt{n+1}$ for the relevant transitions, the qubit remains predominantly in its ground state, such that
\begin{equation}
Q_{g,\ell}\simeq1,
\qquad
Q_{e,\ell}\simeq0.
\label{eq:sm_strong_reset}
\end{equation}
Consequently, the population current reduces to
\begin{equation}
J_{\ell,n}
\simeq
R_{\ell,n}P_n,
\end{equation}
which corresponds to an effectively unidirectional photon-raising process $\ket{n}\rightarrow \ket{n+1}$.

To obtain a finite population model, we truncate the cavity Fock ladder at a finite distance above the target state. 
We denote by $M$ the number of additional Fock states retained above $|N_q\rangle$. 
The reduced population model therefore includes
\begin{equation}
|0\rangle,|1\rangle,\ldots,|N_q\rangle,
|N_q+1\rangle,\ldots,|N_{\rm max}\rangle ,
\end{equation}
with the upper boundary $N_{\rm max}=N_q+M$.
The states above $|N_q\rangle$ account for off-resonant population leakage beyond the target state. 
Increasing $M$ incorporates progressively higher-lying leakage transitions and provides a systematic convergence of the reduced population model. 
In the calculations presented here, we retain the two lowest leakage states by setting $M=2$, such that
\begin{equation}
N_{\rm max}=N_q+2.
\label{eq:sm_Nmax}
\end{equation}

We now construct the upward and downward population-transfer rates of the cavity from the qubit-resolved population currents in Eq.\,\eqref{eq:sm_reduced_current}. 
For the transition $\ket{n}\rightarrow\ket{n+1}$, the forward term contributed by qubit $\ell$ is $Q_{g,\ell}R_{\ell,n}P_n$. 
Since the transition rate is the coefficient multiplying the initial-state population $P_n$, qubit $\ell$ contributes $Q_{g,\ell}R_{\ell,n}$ to the upward population-transfer rate. 
Summing over all auxiliary qubits, we define $\lambda_n$ as the total upward population-transfer rate from $\ket{n}$ to $\ket{n+1}$, namely, the coefficient multiplying $P_n$ in the total forward population current
\begin{equation}
\lambda_n
=
\sum_{\ell=0}^{N_q-1}
Q_{g,\ell}R_{\ell,n},
\qquad
n=0,\ldots,N_{\rm max}-1.
\label{eq:sm_lambda}
\end{equation}

Having obtained the total upward population-transfer rate $\lambda_n$, we now turn to the downward population transfer $\ket{n}\rightarrow\ket{n-1}$. 
The corresponding total downward rate coefficient $\mu_n$ is defined such that the downward population current out of $\ket{n}$ is $\mu_n P_n$. 
It contains two contributions: single-photon cavity loss and reverse AJC population transfer. 
We first evaluate the cavity-loss contribution.
We denote by $\left.\dot P_n\right|_{\kappa}$ the contribution to $\dot P_n$ arising solely from the cavity-loss dissipator $\kappa\mathcal{D}[\hat a]$. 
Projecting this contribution onto the $n$-photon state $\ket{n}$ gives
\begin{align}
\left.\dot P_n\right|_{\kappa}
=
\kappa
\langle n|
\mathcal{D}[\hat a]\hat\rho_c
|n\rangle
=
\kappa(n+1) P_{n+1}
-
\kappa nP_n.
\label{eq:sm_cavity_loss_population}
\end{align}
The first term $\kappa(n+1)P_{n+1}$ describes population transferred into $\ket{n}$ from $\ket{n+1}$ through the loss of one photon, whereas the second term $-\kappa nP_n$ describes population transferred out of $\ket{n}$ toward $\ket{n-1}$. 
Defining $\mu_n^{(\kappa)}$ as the cavity-loss contribution to the total downward rate coefficient $\mu_n$, such that the associated downward population current is $\mu_n^{(\kappa)}P_n$, we obtain
\begin{equation}
\mu_n^{(\kappa)}
=
n\kappa,
\qquad
n=1,\ldots,N_{\rm max}.
\label{eq:sm_cavity_downward_rate}
\end{equation}
The second contribution to $\mu_n$ follows directly from the reverse term in Eq.\,\eqref{eq:sm_reduced_current}. 
For the downward transition $\ket{n}\rightarrow\ket{n-1}$, this equation is evaluated at $n-1$, and the reverse population current contributed by qubit $\ell$ is $Q_{e,\ell}R_{\ell,n-1}P_n$. 
Summing over all auxiliary qubits, we define $\mu_n^{(\mathrm{AJC})}$ as the reverse AJC contribution to the total downward rate coefficient for the transition $\ket{n}\rightarrow\ket{n-1}$, such that the associated downward population current is $\mu_n^{(\mathrm{AJC})}P_n$, with
\begin{equation}
\mu_n^{(\mathrm{AJC})}
=
\sum_{\ell=0}^{N_q-1}
Q_{e,\ell}R_{\ell,n-1},
\qquad
n=1,\ldots,N_{\rm max}.
\label{eq:sm_reverse_AJC_downward_rate}
\end{equation}
Combining this result with the cavity-loss contribution, the total downward rate coefficient is
\begin{align}
\mu_n
=
\mu_n^{(\kappa)}
+
\mu_n^{(\mathrm{AJC})}
=
n\kappa
+
\sum_{\ell=0}^{N_q-1}
Q_{e,\ell}R_{\ell,n-1},
\qquad
n=1,\ldots,N_{\rm max}.
\label{eq:sm_mu}
\end{align}

Taken together, the upward and downward population currents obtained above define a finite nearest-neighbor BD chain for the cavity populations. 
For $n=0,\ldots,N_{\rm max}$, their dynamics is governed by
\begin{equation}
\dot P_n
=
\lambda_{n-1}P_{n-1}
+
\mu_{n+1}P_{n+1}
-
\left(
\lambda_n+\mu_n
\right)P_n,
\label{eq:sm_birth_death}
\end{equation}
with the boundary terms $\lambda_{-1}P_{-1}=0$ and $\mu_{N_{\rm max}+1}P_{N_{\rm max}+1}=0$.
The first two terms describe population flowing into $\ket{n}$ from the neighboring states $\ket{n-1}$ and $\ket{n+1}$, respectively, whereas $\lambda_nP_n$ and $\mu_nP_n$ describe population flowing out of $\ket{n}$ toward $\ket{n+1}$ and $\ket{n-1}$, respectively.
At the two ends of the finite chain, population currents involving states outside the range $0\leq n\leq N_{\rm max}$ are absent, and the outward rate coefficients satisfy
\begin{equation}
\mu_0=0,
\qquad
\lambda_{N_{\rm max}}=0.
\label{eq:sm_birth_death_boundaries}
\end{equation}

Under the strong-relaxation approximation introduced in Eq.\,\eqref{eq:sm_strong_reset}, the auxiliary qubits remain predominantly in their ground states. 
The reverse AJC contribution to the downward population transfer is therefore negligible, whereas cavity loss remains unchanged. 
The upward and downward rates consequently reduce to
\begin{equation}
\lambda_n
=
\sum_{\ell=0}^{N_q-1}R_{\ell,n},
\qquad
\mu_n=n\kappa.
\label{eq:sm_rates_reset}
\end{equation}
Thus, the photon-raising rate is obtained by summing the contributions from all auxiliary qubits, while the photon-lowering rate is determined solely by cavity loss in this approximation.

For uniform couplings, relaxation rates, and qubit pure-dephasing rates, $g_\ell=g$, $\gamma_\ell=\gamma$, and $\gamma_{\phi,\ell}=\gamma_\phi$, respectively, the total decay rate of the coherence $\Gamma_{\ell,n}$ becomes independent of the qubit index $\ell$ and takes the form
\begin{equation}
\Gamma_n
=
\frac{\gamma}{2}
+
\frac{(2n+1)\kappa}{2}
+
2\gamma_\phi
+
\frac{\kappa_\phi}{2}.
\label{eq:sm_Gamma_uniform}
\end{equation}

Substituting the selective detuning obtained above and Eq.\,\eqref{eq:sm_Gamma_uniform} into Eq.\,\eqref{eq:sm_R}, and using Eq.\,\eqref{eq:sm_rates_reset} yields
\begin{equation}
\lambda_n
=
\sum_{\ell=0}^{N_q-1}
\frac{
2g^2(n+1)\Gamma_n
}{
\Gamma_n^2+
\left[
2(n-\ell)K+\delta_{\rm com}
\right]^2
}.
\label{eq:sm_lambda_uniform}
\end{equation}
At $\delta_{\rm com}=0$, the term with $\ell=n$ is resonant for $n=0,\ldots,N_q-1$. 
Hence, each photon-raising transition $|n\rangle\rightarrow|n+1\rangle$ below the target state contains one resonant contribution. 
For $n=N_q$, however, the resonant term $\ell=n$ is absent from the sum, and the smallest detuning magnitude is $|2K|$, attained for $\ell=N_q-1$. 
Consequently, the upward rate $\lambda_{N_q}$ is suppressed relative to the rates below the target, thereby inhibiting population transfer from $|N_q\rangle$ to $|N_q+1\rangle$.

\subsection{Time-dependent solution of the rate equation}
\label{sec:sm_time_solution}

Under the strong-relaxation approximation in Eq.\,\eqref{eq:sm_strong_reset}, the rate coefficients
$\lambda_n$ and $\mu_n$ in Eq.\,\eqref{eq:sm_birth_death} are independent of time.  Equation~\eqref{eq:sm_birth_death} therefore forms a closed linear system for the cavity Fock-state populations with a time-independent generator.
Collecting these populations into the column vector
\begin{equation}
\mathbf P(t)
=
\left[
P_0(t),P_1(t),\ldots,P_{N_{\rm max}}(t)
\right]^T,
\end{equation}
we obtain
\begin{equation}
\frac{d\mathbf P}{dt}
=
\mathbf A\mathbf P,
\label{eq:sm_matrix_rate}
\end{equation}
where $\mathbf A$ is an
$(N_{\rm max}+1)\times(N_{\rm max}+1)$ tridiagonal rate matrix whose nonzero elements are
\begin{equation}
\begin{aligned}
A_{n,n}
&=
-(\lambda_n+\mu_n),
&& n=0,\ldots,N_{\rm max},
\\
A_{n,n-1}
&=
\lambda_{n-1},
&& n=1,\ldots,N_{\rm max},
\\
A_{n,n+1}
&=
\mu_{n+1},
&& n=0,\ldots,N_{\rm max}-1.
\end{aligned}
\label{eq:sm_rate_matrix}
\end{equation}
All other matrix elements vanish.

For an arbitrary initial population vector $\mathbf P(0)$, the time-independent generator $\mathbf A$ gives
\begin{equation}
\mathbf P(t)
=
e^{\mathbf A t}\mathbf P(0).
\label{eq:sm_time_evolution}
\end{equation}
Equation~\eqref{eq:sm_time_evolution} is the exact solution of the finite reduced BD equation defined by Eqs.\,\eqref{eq:sm_matrix_rate} and \eqref{eq:sm_rate_matrix}.
For the dynamics considered here, the cavity is initially in the vacuum state, such that
\begin{equation}
\mathbf P(0)
=
\left[
1,0,\ldots,0
\right]^T.
\label{eq:sm_initial_population}
\end{equation}
The reduced-model populations are evaluated from Eq.\,\eqref{eq:sm_time_evolution} at the same evolution times as the full Lindblad dynamics.

\subsection{Closed-form steady-state solution}
\label{sec:sm_steady_state}

We define the net population current from $\ket{n}$ to $\ket{n+1}$ as
\begin{equation}
J_n
=
\lambda_n P_n
-
\mu_{n+1}P_{n+1},
\qquad
n=0,\ldots,N_{\rm max}-1.
\label{eq:sm_population_current}
\end{equation}
Here, $\lambda_nP_n$ describes the upward population transfer from $\ket{n}$ to $\ket{n+1}$, whereas $\mu_{n+1}P_{n+1}$ describes the downward population transfer from $\ket{n+1}$ to $\ket{n}$. 
Consequently, $J_n>0$ ($J_n<0$) indicates net population transfer toward higher (lower) photon numbers.
With this definition, Eq.\,\eqref{eq:sm_birth_death} can be written in the continuity form
\begin{equation}
\dot P_n
=
J_{n-1}-J_n,
\qquad
n=0,\ldots,N_{\rm max},
\label{eq:sm_continuity}
\end{equation}
where $J_{-1}=J_{N_{\rm max}}=0$ specifies the closed boundary conditions.

The steady-state population distribution is obtained by imposing $\dot P_n=0$ in Eq.\,\eqref{eq:sm_birth_death}. 
We denote the steady-state populations by $P_n^{\rm ss}$ and the corresponding population current by
\begin{equation}
J_n^{\rm ss}
\equiv
\lambda_n P_n^{\rm ss}
-
\mu_{n+1}P_{n+1}^{\rm ss},
\qquad
n=0,\ldots,N_{\rm max}-1.
\label{eq:sm_steady_population_current}
\end{equation}
The steady-state form of Eq.\,\eqref{eq:sm_continuity} is therefore
\begin{equation}
0
=
J_{n-1}^{\rm ss}
-
J_n^{\rm ss}.
\label{eq:sm_stationary_continuity}
\end{equation}
For $n=0$, together with the boundary condition $J_{-1}=0$, Eq.\,\eqref{eq:sm_stationary_continuity} gives
\begin{equation}
J_0^{\rm ss}=0.
\end{equation}
Applying Eq.\,\eqref{eq:sm_stationary_continuity} successively for $n=1,\ldots,N_{\rm max}-1$ then gives
\begin{equation}
J_n^{\rm ss}=0,
\qquad
n=0,\ldots,N_{\rm max}-1.
\label{eq:sm_zero_current}
\end{equation}
Using the definition of $J_n^{\rm ss}$, the steady-state populations therefore satisfy
\begin{equation}
\lambda_n P_n^{\rm ss}
=
\mu_{n+1}P_{n+1}^{\rm ss},
\qquad
n=0,\ldots,N_{\rm max}-1.
\label{eq:sm_stationary_balance}
\end{equation}

Using the local balance relation in Eq.\,\eqref{eq:sm_stationary_balance}, we introduce the ratio of adjacent steady-state populations,
\begin{equation}
r_n
\equiv
\frac{P_{n+1}^{\rm ss}}{P_n^{\rm ss}}
=
\frac{\lambda_n}{\mu_{n+1}},
\qquad
n=0,\ldots,N_{\rm max}-1.
\label{eq:sm_population_ratio}
\end{equation}
The ratio $r_n$ compares the effective upward rate from $\ket{n}$ to $\ket{n+1}$ with the corresponding downward rate from $\ket{n+1}$ to $\ket{n}$.

Iterating Eq.\,\eqref{eq:sm_population_ratio} from the vacuum population gives
\begin{equation}
P_n^{\rm ss}
=
P_0^{\rm ss}
\prod_{k=0}^{n-1}r_k,
\qquad
n=1,\ldots,N_{\rm max}.
\label{eq:sm_steady_product}
\end{equation}
Imposing the normalization condition
\begin{equation}
\sum_{n=0}^{N_{\rm max}}P_n^{\rm ss}=1
\end{equation}
then yields
\begin{equation}
P_n^{\rm ss}
=
\frac{
\displaystyle
\prod_{k=0}^{n-1}r_k
}{
\displaystyle
\sum_{j=0}^{N_{\rm max}}
\prod_{k=0}^{j-1}r_k
},
\qquad
n=0,\ldots,N_{\rm max},
\label{eq:sm_steady_solution}
\end{equation}
where an empty product is defined as unity. 
Substituting $r_k=\lambda_k/\mu_{k+1}$ gives the steady-state distribution directly in terms of the birth and death rates.

\subsection{Stabilization criteria for the target Fock state}
\label{sec:sm_target_condition}

Equation~\eqref{eq:sm_population_ratio} directly determines whether the steady-state population increases or decreases between adjacent Fock states. 
Specifically, we have
\begin{equation}
\begin{aligned}
r_n>1
&\quad\Longleftrightarrow\quad
P_{n+1}^{\rm ss}>P_n^{\rm ss},
\\
r_n<1
&\quad\Longleftrightarrow\quad
P_{n+1}^{\rm ss}<P_n^{\rm ss}.
\end{aligned}
\label{eq:sm_ratio_interpretation}
\end{equation}
A strict local maximum of the population distribution at the target state $\ket{N_q}$ is equivalent to
\begin{equation}
r_{N_q-1}>1,
\qquad
r_{N_q}<1.
\label{eq:sm_local_target_condition}
\end{equation}
These two inequalities compare the target population only with its two nearest neighbors.
To relate the population-ratio sequence to the total target-state population, we rewrite the steady-state distribution in Eq.\,\eqref{eq:sm_steady_solution} relative to $P_{N_q}^{\rm ss}$. 
This gives
\begin{equation}
P_{N_q}^{\rm ss}
=
\left[
1
+
\sum_{j=0}^{N_q-1}
\left(
\prod_{k=j}^{N_q-1}r_k
\right)^{-1}
+
\sum_{j=N_q+1}^{N_{\rm max}}
\left(\prod_{k=N_q}^{j-1}r_k\right)
\right]^{-1}.
\label{eq:sm_target_population}
\end{equation}
The first sum is the total population below the target relative to $P_{N_q}^{\rm ss}$, whereas the second is the corresponding relative population above the target. 
Therefore, the two adjacent conditions in Eq.\,\eqref{eq:sm_local_target_condition} establish a local population maximum but do not by themselves determine the total population carried by the target state.
A simplified sufficient, although not necessary, regime for a steady-state distribution sharply concentrated at $\ket{N_q}$ is
\begin{equation}
\begin{aligned}
r_n&\gg1,
&& n=0,\ldots,N_q-1,
\\
r_n&\ll1,
&& n=N_q,\ldots,N_{\rm max}-1.
\end{aligned}
\label{eq:sm_target_condition}
\end{equation}
Under this hierarchy, every contribution to the two sums in Eq.\,\eqref{eq:sm_target_population} is suppressed.

The population-ratio hierarchy in Eq.\,\eqref{eq:sm_target_condition} can be made explicit directly from Eq.\,\eqref{eq:sm_lambda_uniform}.
We set $\delta_{\rm com}=0$ and consider the resolved-transition regime $2|K|\gg\Gamma_n$. 
In this regime, $\lambda_n$ is dominated by the resonant contribution from qubit $\ell=n$, yielding
\begin{equation}
\lambda_n
\simeq
R_{n,n}
=
\frac{2g^2(n+1)}{\Gamma_n},
\qquad
n=0,\ldots,N_q-1.
\label{eq:sm_lambda_resonant}
\end{equation}
Using $\mu_{n+1}=(n+1)\kappa$ from Eq.\,\eqref{eq:sm_rates_reset}, the corresponding population
ratios are
\begin{equation}
r_n=\frac{\lambda_n}{\mu_{n+1}}
\simeq
\frac{2g^2}{\kappa\Gamma_n},
\qquad
n=0,\ldots,N_q-1.
\label{eq:sm_ratio_below_target}
\end{equation}
Hence, in the strong-relaxation regime, the resonant AJC channel produces net upward population transfer provided that
\begin{equation}
\frac{2g^2}{\kappa\Gamma_n}\gg1,
\qquad
n=0,\ldots,N_q-1.
\label{eq:sm_condition_below_target}
\end{equation}
For $n\geq N_q$, none of the auxiliary qubits is resonant with the transition $\ket{n}\leftrightarrow\ket{n+1}$. 
Substituting $\delta_{\ell,n}=2(n-\ell)K$ into Eq.\,\eqref{eq:sm_lambda_uniform} gives
\begin{equation}
\lambda_n
=
\sum_{\ell=0}^{N_q-1}
\frac{
2g^2(n+1)\Gamma_n
}{
\Gamma_n^2+4(n-\ell)^2K^2
},
\qquad
n=N_q,\ldots,N_{\rm max}-1.
\label{eq:sm_lambda_above_target}
\end{equation}
In the resolved-transition regime $2|K|\gg\Gamma_n$, this becomes
\begin{equation}
\lambda_n
\simeq
\frac{g^2(n+1)\Gamma_n}{2K^2}
\sum_{\ell=0}^{N_q-1}
\frac{1}{(n-\ell)^2}.
\label{eq:sm_lambda_above_target_largeK}
\end{equation}
Using $\mu_{n+1}=(n+1)\kappa$, the corresponding population ratios are
\begin{equation}
r_n
\simeq
\frac{g^2\Gamma_n}{2\kappa K^2}
\sum_{\ell=0}^{N_q-1}
\frac{1}{(n-\ell)^2},
\qquad
n=N_q,\ldots,N_{\rm max}-1.
\label{eq:sm_ratio_above_target}
\end{equation}
Hence, a simplified sufficient condition for suppressing the complete upper population tail is
\begin{equation}
\frac{g^2\Gamma_n}{2\kappa K^2}
\sum_{\ell=0}^{N_q-1}
\frac{1}{(n-\ell)^2}
\ll1,
\qquad
n=N_q,\ldots,N_{\rm max}-1.
\label{eq:sm_condition_above_target}
\end{equation}
The two conditions determining the nearest-neighbor contrast around the target state are obtained by taking $n=N_q-1$ in Eq.\,\eqref{eq:sm_condition_below_target} and $n=N_q$ in Eq.\,\eqref{eq:sm_condition_above_target}
\begin{equation}
\frac{2g^2}{\kappa\Gamma_{N_q-1}}\gg1,
\qquad
\frac{g^2\Gamma_{N_q}}{2\kappa K^2}
\sum_{\ell=0}^{N_q-1}
\frac{1}{(N_q-\ell)^2}
\ll1.
\label{eq:sm_local_parameter_conditions}
\end{equation}
The first condition suppresses the population of $\ket{N_q-1}$ relative to $\ket{N_q}$, whereas the second suppresses the population of $\ket{N_q+1}$ relative to $\ket{N_q}$. They therefore establish the nearest-neighbor conditions for a local population maximum. 
The total target-state population is instead determined by the complete ratio hierarchy in Eqs.\,\eqref{eq:sm_condition_below_target} and \eqref{eq:sm_condition_above_target}, or equivalently by the two cumulative tail contributions in Eq.\,\eqref{eq:sm_target_population}.

The full hierarchy in Eqs.\,\eqref{eq:sm_condition_below_target} and \eqref{eq:sm_condition_above_target} provides a simplified sufficient operating regime for concentrating the steady-state distribution near the target Fock state.
A sufficiently large Kerr nonlinearity $K$ suppresses the off-resonant transition above the target, whereas pure dephasing in the resolved-transition regime should remain weak because its contribution to $\Gamma_n$ simultaneously reduces the resonant population transfer below the target and enhances the residual leakage above it. 
The cavity-loss rate $\kappa$, however, should not simply be minimized. 
Instead, it must lie in an intermediate window: it should be weak enough not to overwhelm the resonant AJC transfer into the target state, but strong enough to remove population generated by the residual off-resonant transition out of it. 
Qubit relaxation likewise plays a dual role: it must be sufficiently strong to complete the irreversible relaxation, while its contribution to $\Gamma_n$ must remain small enough to maintain strong resonant transfer below the target and weak off-resonant leakage above it.

For the cascaded configuration of the main text, qubits $\ell=0,\ldots,N_q-1$ resonantly address the successive transitions $|0\rangle\rightarrow|1\rangle\rightarrow\cdots\rightarrow|N_q\rangle$.
Accordingly, the resonant terms dominate $\lambda_n$ for $n<N_q$, whereas for $n\geq N_q$ all contributions to $\lambda_n$ are off-resonant. 
In the resolved-transition regime, this produces a strong upward bias below the target while suppressing further excitation beyond $|N_q\rangle$, yielding a steady-state distribution concentrated near the target Fock state.

The time-dependent solution in Eq.\,\eqref{eq:sm_time_evolution} and the steady-state solution in Eq.\,\eqref{eq:sm_steady_solution} are exact solutions of the reduced BD equation. 
The approximations enter in deriving this equation from the full Lindblad dynamics: we retain only the qubit-resolved transition coherences $C_{\ell,n}$, neglect their coupling to neighboring transition coherences and to correlations generated by the other auxiliary qubits, adiabatically eliminate the retained coherences, and factorize the joint populations as $P_{n,g_\ell}\simeq P_n Q_{g,\ell}$ and $P_{n,e_\ell}\simeq P_n Q_{e,\ell}$. 
The closed-form analytical results in the strong-relaxation regime additionally use $Q_{g,\ell}\simeq1$ and $Q_{e,\ell}\simeq0$. 
For the moderate relaxation rate $\gamma/g=3$ used in the numerical comparisons, this approximation is not asymptotically controlled for the highest appreciably populated Fock states. 
Its quantitative accuracy, together with that of the approximations introduced above, is therefore assessed directly by comparison with the full Lindblad dynamics.
Apart from finite-cutoff errors, discrepancies between the two results therefore quantify the approximations introduced in reducing the full Lindblad dynamics to the BD equation.
The cavity Hilbert-space truncation in the full Lindblad calculation and the population cutoff $N_{\rm max}$ in the reduced model are distinct numerical approximations and should be checked independently for convergence.

\subsection{Convergence with the full Lindblad cavity Hilbert-space truncation}
\label{sec:sm_cavity_truncation}

In the full Lindblad simulations presented in this work, we use a cavity Hilbert-space dimension of $N_{\rm cav}=12$, corresponding to the retained Fock states $n=0,\ldots,11$.
To verify the convergence of this choice, we independently recompute the steady state for $N_q=4$ using $N_{\rm cav}=6,8,10,$ and $12$.
Here $N_{\rm cav}$ denotes the number of retained cavity Fock states. 
The four steady-state quantities reported are the target-state population $P^{\rm ss}_4$, the mean photon number $\langle n\rangle_{\rm ss}$, the integrated Wigner negativity $\mathcal N_W$, and the population $P^{\rm ss}_{N_{\rm cav}-1}$ of the highest retained Fock state, with the last quantity directly probing accumulation at the artificial truncation boundary.

\begin{table}[t]
\caption{Convergence of the full Lindblad steady-state calculation with respect to the cavity Hilbert-space truncation $N_{\rm cav}$ for $N_q=4$. 
The cavity Fock basis is truncated to $n=0,\ldots,N_{\rm cav}-1$. 
The parameters are $K/g=100$, $\kappa/g=0.05$, $\gamma/g=3$, $\delta_{\rm com}/g=0$, and $\gamma_{\phi}/g=\kappa_{\phi}/g=0$.
}
\label{tab:cavity_truncation}
\centering
\begin{ruledtabular}
\begin{tabular}{ccccc}
$N_{\rm cav}$
& $P^{\rm ss}_4$
& $\langle n\rangle_{\rm ss}$
& $\mathcal{N}_W$
& $P^{\rm ss}_{N_{\rm cav}-1}$ \\
\hline
6  & 0.8931850619 & 3.8869967021 & 0.49355345 & $2.058532\times10^{-3}$ \\
8  & 0.8931836012 & 3.8870000319 & 0.49355245 & $5.766555\times10^{-10}$ \\
10 & 0.8931836012 & 3.8870000319 & 0.49355245 & $<10^{-14}$ \\
12 & 0.8931836012 & 3.8870000319 & 0.49355245 & $<10^{-14}$ \\
\end{tabular}
\end{ruledtabular}
\end{table}

As shown in Table~\ref{tab:cavity_truncation}, the values of $P^{\rm ss}_4$, $\langle n\rangle_{\rm ss}$, and $\mathcal{N}_W$ are already unchanged to the reported precision for $N_{\rm cav}\geq 8$, while the population at the truncation boundary rapidly decreases from $2.06\times10^{-3}$ at $N_{\rm cav}=6$ to below $10^{-14}$ for $N_{\rm cav}\geq 10$. 
These results confirm that $N_{\rm cav}=12$, as used in the full Lindblad simulations, is safely within the converged regime and that the reported steady-state observables are free of appreciable cavity-cutoff artifacts.

\subsection{Convergence with the population cutoff}
\label{sec:sm_cutoff_convergence}

\begin{figure}[htb]
\centerline{\includegraphics[width=9cm]{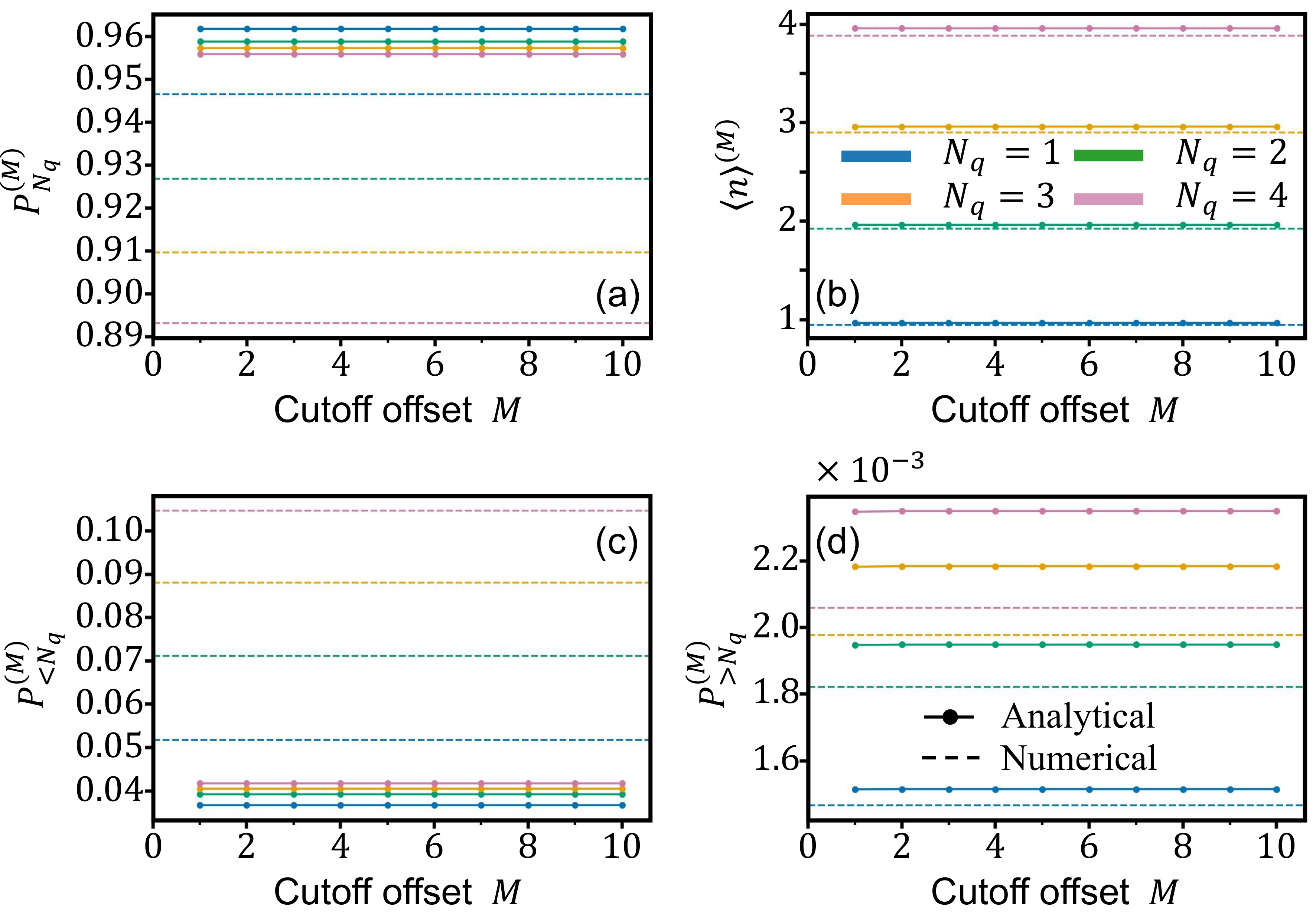}}
\caption{Convergence of the analytical steady-state solution with the cutoff offset $M$ for target Fock states $N_q=1$--$4$.
The analytical BD model retains Fock states up to $N_{\rm max}=N_q+M$.
Shown are (a) the target-state population $P_{N_q}^{(M)}$, (b) the mean photon number $\langle n\rangle^{(M)}$, (c) the total population below the target, $P_{<N_q}^{(M)}=\sum_{n=0}^{N_q-1}P_n^{(M)}$, and (d) the total population above the target, $P_{>N_q}^{(M)}=\sum_{n=N_q+1}^{N_q+M}P_n^{(M)}$.
Solid lines with markers denote the analytical BD results, while the horizontal broken lines show the corresponding full Lindblad numerical benchmarks.
The weak dependence on $M$ demonstrates rapid convergence of the analytical truncation.
Parameters are $K/g=100$, $\gamma/g=3$, $\kappa/g=0.05$, $\delta_{\rm com}/g=0$, and $\gamma_\phi/g=\kappa_\phi/g=0$.
}
\label{figS1}
\end{figure}

We next examine the convergence of the analytical steady-state solution with respect to the finite population cutoff introduced above.
Recall that $M$ denotes the number of additional Fock states retained above the target state, such that $N_{\rm max}=N_q+M$.
For each target Fock state $N_q$, we denote by $P_n^{(M)}$ the normalized steady-state population obtained from the truncated BD model with cutoff offset $M$.

To monitor the convergence of both the target population and the population tails, we consider the target-state population $P_{N_q}^{(M)}$ together with the following three quantities
\begin{align}
\langle n\rangle^{(M)}
&=
\sum_{n=0}^{N_q+M}
nP_n^{(M)},
\\
P_{<N_q}^{(M)}
&=
\sum_{n=0}^{N_q-1}
P_n^{(M)},
\\
P_{>N_q}^{(M)}
&=
\sum_{n=N_q+1}^{N_q+M}
P_n^{(M)}.
\end{align}
Here, $\langle n\rangle^{(M)}$ is the mean cavity photon number, i.e., the first moment of the cavity Fock-state distribution, while $P_{<N_q}^{(M)}$ and $P_{>N_q}^{(M)}$ are the total populations below and above the target state, respectively.
The three contributions form a complete partition of the normalized steady-state population and therefore satisfy
\begin{equation}
P_{<{N_q}}^{(M)}
+
P_{N_q}^{(M)}
+
P_{>{N_q}}^{(M)}
=
1.
\end{equation}

Figure~\ref{figS1} shows these quantities as functions of the cutoff offset $M$ for the target states $N_q=1,\ldots,4$.
The analytical results exhibit an extremely weak dependence on $M$ over the entire range shown.
In particular, both the target-state population $P_{N_q}^{(M)}$ and the mean photon number $\langle n\rangle^{(M)}$ are essentially unchanged as additional Fock states are included above the target.
The same rapid convergence is observed separately in the lower and upper population tails.

The physical origin of this rapid convergence is especially transparent from $P_{>N_q}^{(M)}$.
For the parameters considered here, the total population above the target remains at the $10^{-3}$ level, reflecting the strong suppression of off-resonant photon-raising processes beyond $\ket{N_q}$ by the Kerr-induced spectral selectivity.
Consequently, progressively higher Fock states carry only a negligible fraction of the steady-state population, and extending the analytical cutoff produces only very small corrections to the normalized distribution.
The population below the target is substantially larger than the upper leakage tail, but it is insensitive to the number of additional states retained above $\ket{N_q}$ and is likewise rapidly converged.
Moreover, the residual analytical--numerical discrepancy is concentrated primarily in the population below the target, whereas the upper leakage tail remains small in absolute magnitude.

The horizontal numerical references in Fig.\,\ref{figS1} are obtained from the full Lindblad steady state and are independent of the analytical cutoff $M$.
Importantly, increasing $M$ does not appreciably reduce the remaining difference between the analytical and full Lindblad results.
The residual analytical--numerical discrepancy therefore does not originate from the finite population cutoff, but instead reflects the approximations involved in reducing the full cavity--qubit dynamics to the analytical BD model, as discussed above.
The convergence shown here confirms that retaining $M=2$ additional Fock states, i.e.,
$N_{\rm max}=N_q+2$, is numerically sufficient for the parameter regime investigated here.

\section{Effects of qubit and cavity pure dephasing}
\label{sec:sm_dephasing}

\begin{figure}[htb]
\centerline{\includegraphics[width=9cm]{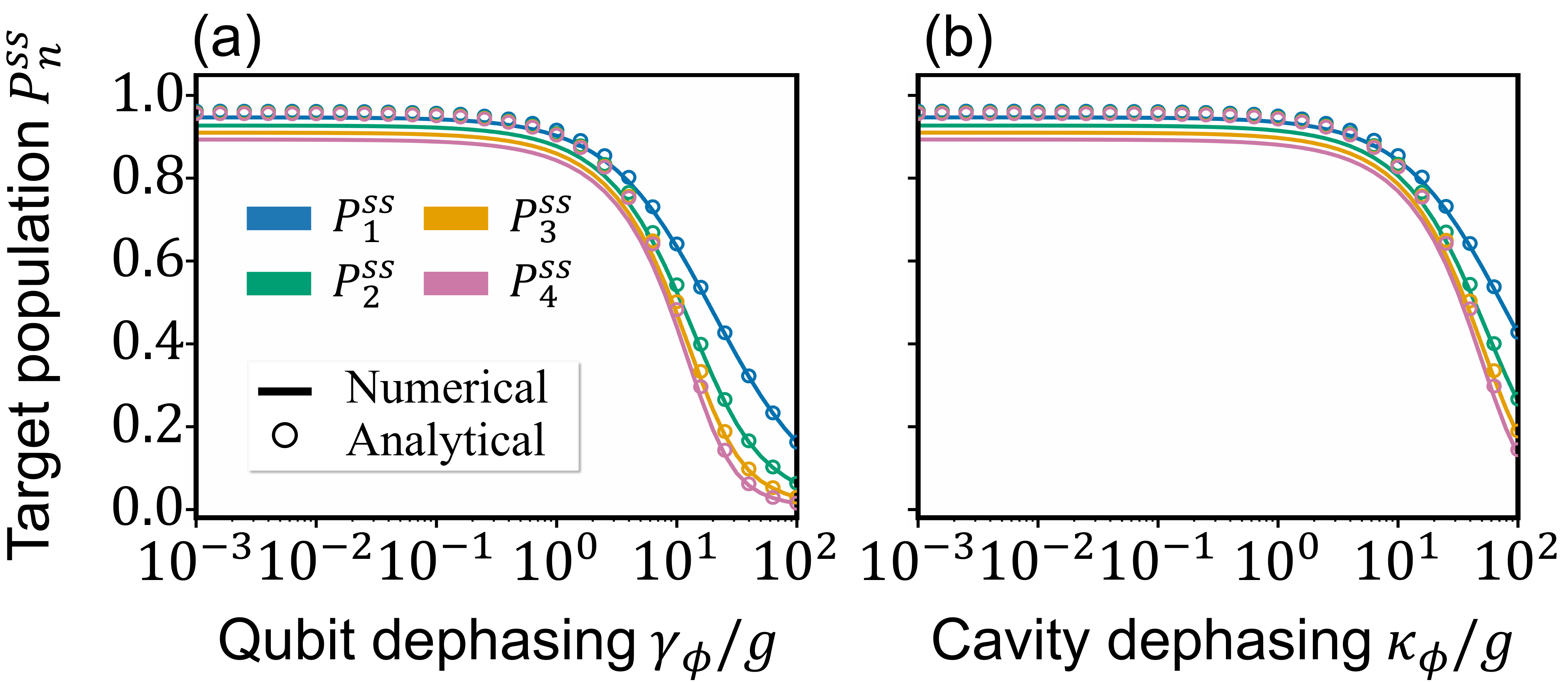}}
\caption{Robustness of the stabilized target Fock-state populations against pure
dephasing. Steady-state target populations $P_{N_q}^{\mathrm{ss}}$
($N_q=1,\ldots,4$) versus (a) the common qubit dephasing parameter
$\gamma_{\phi}/g$, taken to be identical for all auxiliary qubits,
and (b) the cavity dephasing parameter $\kappa_{\phi}/g$.
Solid curves and open circles denote the full numerical Lindblad and analytical BD results, respectively. The unscanned dephasing parameter
is set to zero. Other parameters are $K/g=100$, $\kappa/g=0.05$,
$\gamma/g=3$, and $\delta_{\mathrm{com}}/g=0$.
}
\label{figS2}
\end{figure}

We next examine the influence of qubit and cavity pure dephasing on the stabilized target Fock-state populations. 
In the qubit-dephasing scan, all auxiliary qubits are assigned the same dephasing parameter,
\begin{equation}
\gamma_{\phi,0}
=
\gamma_{\phi,1}
=
\cdots
=
\gamma_{\phi,N_q-1}
\equiv
\gamma_\phi .
\label{eq:sm_common_qubit_dephasing}
\end{equation}
As mentioned above, qubit and cavity pure dephasing enter the reduced population dynamics through the total decay rate of the AJC transition coherences.
For uniform parameters, the total decay rate of the coherence can be written as
\begin{equation}
\Gamma_n
=
\Gamma_n^{(0)}
+
2\gamma_\phi
+
\frac{\kappa_\phi}{2},
\qquad
\Gamma_n^{(0)}
=
\frac{\gamma}{2}
+
\frac{(2n+1)\kappa}{2}.
\label{eq:sm_Gamma_dephasing}
\end{equation}
Thus, within the reduced BD description, both dephasing parameters affect the reduced cavity population dynamics through their contributions to the total coherence decay rate $\Gamma_n$, which enters the effective exchange coefficient $R_{\ell,n}$ defined in Eq.\,\eqref{eq:sm_R}.
The different prefactors of $\gamma_\phi$ and $\kappa_\phi$ follow from the dissipator convention used in the master equation.

At the selective working point $\delta_{\rm com}=0$, the AJC transition $\ket{n,g_n}\leftrightarrow\ket{n+1,e_n}$ is resonant for each $n=0,\ldots,N_q-1$. 
Setting $\ell=n$ and $\delta_{n,n}=0$ in Eq.\,\eqref{eq:sm_R} gives
\begin{equation}
R_{n,n}^{\rm res}
=
\frac{2g^2(n+1)}{\Gamma_n}.
\label{eq:sm_dephasing_resonant}
\end{equation}
Thus, increasing either dephasing rate reduces the resonant contribution $R_{n,n}^{\rm res}$ to the upward rate below the target.
For the transition out of the target state and all higher transitions, $n\geq N_q$, none of the auxiliary qubits is resonant. 
At $\delta_{\rm com}=0$, their detunings are $\delta_{\ell,n}=2(n-\ell)K$, with $\ell=0,\ldots,N_q-1$.
Substituting this relation into Eq.\,\eqref{eq:sm_R}, and assuming $\Gamma_n\ll 2|n-\ell||K|$, gives
\begin{equation}
R_{\ell,n}^{\rm off}
\simeq
\frac{
g^2(n+1)\Gamma_n
}{
2(n-\ell)^2K^2
},
\qquad
\ell\neq n.
\label{eq:sm_dephasing_off_resonant}
\end{equation}
Within this off-resonant regime, increasing either dephasing rate increases the off-resonant contributions $R_{\ell,n}^{\rm off}$ to the upward rates at and above the target.
Pure dephasing therefore reduces the contrast between the resonant contributions to the upward rates below the target and the off-resonant contributions at and above the target, thereby lowering the steady-state target population $P_{N_q}^{\rm ss}$.

Figure~\ref{figS2} shows the resulting steady-state target populations for the stabilized states $\ket{N_q}$, with $N_q=1,\ldots,4$. 
Increasing either the common qubit dephasing parameter $\gamma_\phi$ or the cavity dephasing parameter $\kappa_\phi$ reduces the target-state population. 
The analytical BD results closely follow the full Lindblad calculations, showing that the dephasing dependence is captured primarily by the contributions of the two dephasing channels to the total coherence decay rate.
Since $\Gamma_n$ increases with dephasing, the off-resonant condition $\Gamma_n\ll 2|n-\ell||K|$ may no longer be satisfied at large dephasing rates. Equation~\eqref{eq:sm_dephasing_off_resonant} is therefore used only to explain the behavior in the weak-dephasing regime. 
The analytical curves in Fig.\,\ref{figS2} are evaluated using the full Lorentzian coefficient $R_{\ell,n}$ defined in Eq.\,\eqref{eq:sm_R}, without making the off-resonant approximation.

\section{Stabilization with inhomogeneous auxiliary-qubit parameters}
\label{sec:sm_parameter_inhomogeneity}

In the main text, uniform coupling strengths and relaxation rates are adopted for simplicity and to provide a convenient reference setting. 
Here we relax these simplifying assumptions and examine more general qubit-dependent parameter configurations.
In realistic implementations, the coupling strengths and relaxation rates may vary from qubit to qubit, while the individual qubit detunings may deviate independently from the values required by their assigned resonance conditions.
We therefore investigate the steady-state stabilization under random qubit-to-qubit parameter variations. 
We focus on the $N_q=4$ case, corresponding to the highest target Fock state considered in this work.

We consider independently the inhomogeneity of the AJC coupling strengths and qubit relaxation rates, together with qubit-specific detuning errors.
For the coupling-strength inhomogeneity, the coupling of the first auxiliary qubit is kept fixed as the reference scale, $g_0=g$, while the remaining couplings are independently sampled according to
\begin{equation}
g_\ell
=
g\left(1+\eta_g\xi_\ell^{(g)}\right),
\qquad
\ell=1,2,3,
\label{eq:sm_g_inhomogeneity}
\end{equation}
where $\eta_g$ specifies the relative inhomogeneity strength and the random variables $\xi_\ell^{(g)}$ are independently sampled from the uniform distribution $\mathcal{U}[-1,1]$. 
Thus, for a given value of $\eta_g$, each varied coupling independently lies within the interval $g(1-\eta_g)\leq g_\ell\leq g(1+\eta_g)$.

For the qubit-relaxation-rate inhomogeneity, all four auxiliary-qubit relaxation rates are allowed to vary independently according to
\begin{equation}
\gamma_\ell
=
\gamma\left(1+\eta_\gamma\xi_\ell^{(\gamma)}\right),
\qquad
\ell=0,1,2,3,
\label{eq:sm_gamma_disorder}
\end{equation}
where $\eta_\gamma$ specifies the relative inhomogeneity strength and the random variables $\xi_\ell^{(\gamma)}$ are independently sampled from the uniform distribution $\mathcal{U}[-1,1]$.

To characterize qubit-specific resonance errors with $\delta_{\rm com}=0$, we write the detuning of each auxiliary qubit as
\begin{equation}
\Delta_\ell
=
-(2\ell+1)K+\epsilon_\ell,
\qquad
\ell=0,1,2,3,
\label{eq:sm_detuning_disorder}
\end{equation}
with
\begin{equation}
\frac{\epsilon_\ell}{g}
=
\eta_\Delta\xi_\ell^{(\Delta)}.
\label{eq:sm_detuning_disorder_strength}
\end{equation}
Here $\eta_\Delta$ specifies the dimensionless detuning-error strength, and the random variables $\xi_\ell^{(\Delta)}$ are independently sampled from $\mathcal{U}[-1,1]$.
These qubit-specific detuning errors are distinct from the common detuning $\delta_{\rm com}$ considered in the main text: $\delta_{\rm com}$ shifts all AJC transitions by the same amount, whereas the random offsets $\epsilon_\ell$ independently displace the auxiliary qubits from their respective resonance conditions.

\begin{figure}[htb]
\centerline{\includegraphics[width=9cm]{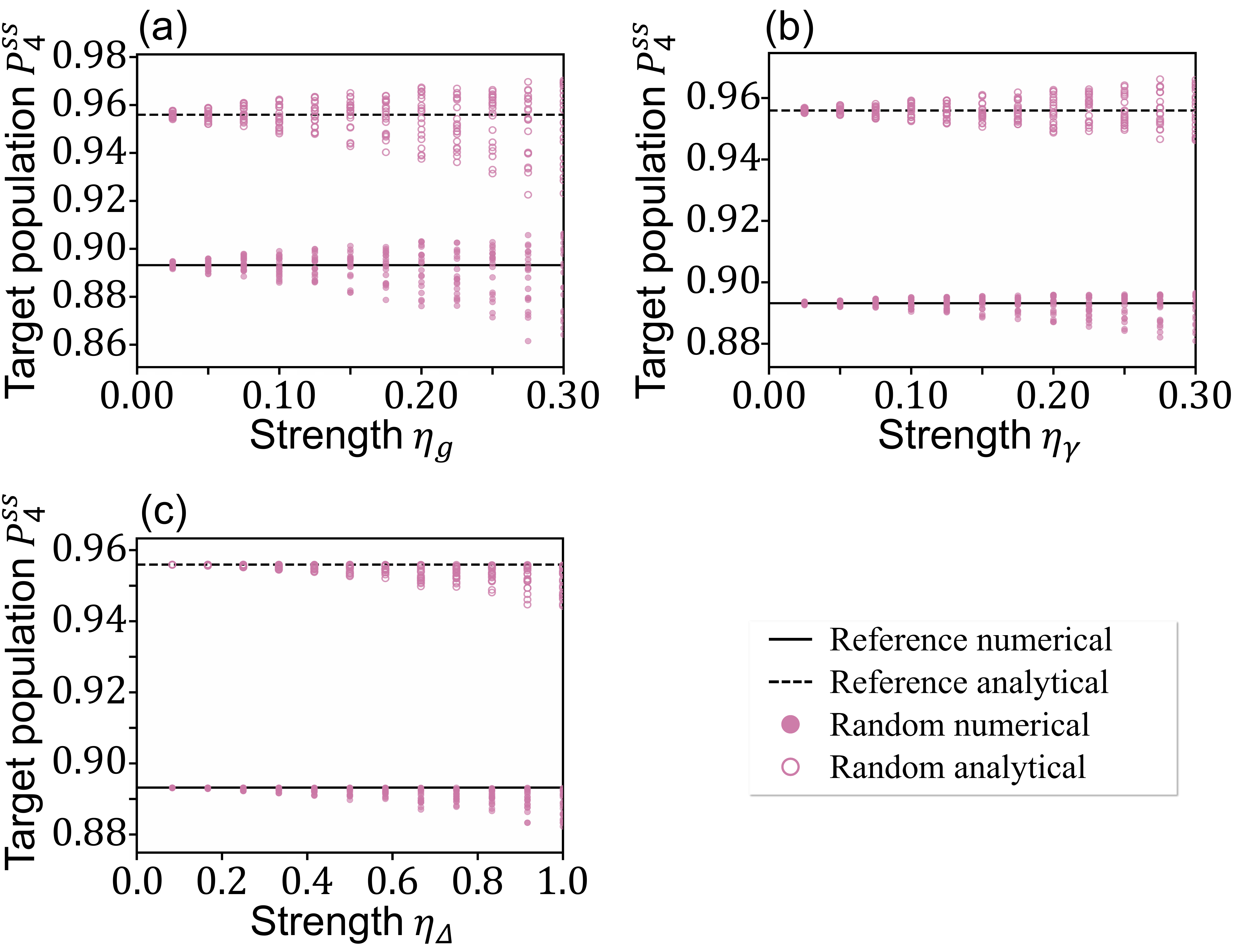}}
\caption{Steady-state stabilization of the target state $|4\rangle$ in the presence of auxiliary-qubit parameter inhomogeneity for $N_q=4$.
The three panels show independent random variations of (a) the AJC coupling strengths $g_\ell$, (b) the qubit relaxation rates $\gamma_\ell$, and (c) the qubit-specific detuning errors $\epsilon_\ell$.
In each panel, only the indicated qubit parameter is varied, while the remaining parameters are kept at their reference values.
For each nonzero value of $\eta_g$, $\eta_\gamma$, or $\eta_\Delta$, $20$ independent random realizations are generated.
For the coupling-strength inhomogeneity in (a), $g_0=g$ is kept fixed as the reference scale, while $g_\ell$ with $\ell=1,2,3$ are independently sampled according to Eq.\,\eqref{eq:sm_g_inhomogeneity}.
For the relaxation-rate inhomogeneity in (b), all four $\gamma_\ell$ are independently sampled according to Eq.\,\eqref{eq:sm_gamma_disorder}, while in (c) the qubit-specific detuning errors $\epsilon_\ell$ are independently sampled according to Eq.\,\eqref{eq:sm_detuning_disorder_strength} and enter the detunings through Eq.\,\eqref{eq:sm_detuning_disorder}.
Filled and open circles show the full numerical Lindblad and analytical BD results, respectively, for the same random realizations.
The horizontal solid and dashed black lines denote the corresponding reference full numerical Lindblad and analytical results.
The reference configuration is $g_\ell=g$, $\gamma_\ell=\gamma$, and $\epsilon_\ell=0$, with $K/g=100$, $\kappa/g=0.05$, $\gamma/g=3$, $\delta_{\rm com}/g=0$, and $\gamma_\phi/g=\kappa_\phi/g=0$.
}
\label{figS3}
\end{figure}

For each value of $\eta_g$, $\eta_\gamma$, or $\eta_\Delta$, multiple realizations are generated by independently sampling the corresponding qubit-dependent parameters according to the distributions defined above.
For each realization, the steady-state target population $P_4^{\rm ss}$ is evaluated from the full Lindblad master equation.
The same randomly generated parameter set is also used in the analytical BD model, so that the full Lindblad and analytical results are compared using exactly the same qubit-dependent parameters.
Figure~\ref{figS3} shows the reference full Lindblad and analytical results as horizontal lines and the results for the individual random realizations as scatter points.
The vertical spread of the points at a fixed inhomogeneity strength therefore directly shows how $P_4^{\rm ss}$ varies among the independently generated parameter sets.

As shown in Fig.\,\ref{figS3}, the reference setting with uniform coupling strengths and relaxation rates serves as a convenient baseline rather than an optimized parameter configuration. 
For the coupling-strength and relaxation-rate variations in Figs.\,\ref{figS3}(a) and \ref{figS3}(b), respectively, different randomly generated parameter sets can yield $P_4^{\rm ss}$ either below or above the corresponding reference value. 
This shows that departures from uniform $g_\ell$ or $\gamma_\ell$ do not necessarily reduce the stabilization performance. 
As $\eta_g$ or $\eta_\gamma$ increases, the spread of $P_4^{\rm ss}$ among the sampled parameter sets becomes larger, with the effect being more pronounced for the coupling-strength variations.
In contrast, the qubit-specific detuning errors in Fig.\,\ref{figS3}(c) shift the auxiliary qubits away from their assigned AJC resonance conditions and therefore predominantly reduce $P_4^{\rm ss}$ as $\eta_\Delta$ increases. 
The analytical BD results reproduce these qualitative trends, although they systematically predict higher absolute target-state populations than the full Lindblad calculations.

Overall, for the $N_q=4$ case and over the parameter ranges considered, these results demonstrate that effective stabilization does not rely on uniform auxiliary-qubit coupling strengths or relaxation rates.
The reference choice $g_\ell=g$ and $\gamma_\ell=\gamma$ therefore serves as a convenient benchmark rather than an optimized configuration, whereas maintaining the assigned AJC resonance conditions remains important for preserving a high target-state population.
More generally, what matters is not whether the auxiliary-qubit parameters are identical, but whether the qubit-dependent parameters satisfy the conditions required by the stabilization mechanism discussed above.

\section{Experimental implementation and feasibility}
\label{sec:sm_experimental_feasibility}

The effective model in Eq.\,\eqref{eq:sm_H} can be implemented in a superconducting circuit quantum electrodynamics (QED) architecture comprising a Kerr-nonlinear cavity coupled to $N_q$ auxiliary qubits, together with engineered qubit-relaxation channels and individually applied blue-sideband modulation drives. 
The required ingredients, including Kerr nonlinearity, controlled qubit relaxation, and parametrically activated sideband transitions, have each been demonstrated experimentally~\cite{Kirchmair2013,Reed2010APL,Strand2013,Leek2009}.
The cavity Kerr nonlinearity can be engineered by incorporating a Josephson nonlinear element into the resonator, such as a superconducting quantum interference device (SQUID)~\cite{Tancredi2013}, a superconducting nonlinear asymmetric inductive element (SNAIL)~\cite{FrattiniAPL2017,Lu2023SNAIL}, or a Josephson-junction array~\cite{Weissl2015}. 
A recent experiment using a SNAIL-terminated superconducting resonator demonstrated a flux-tunable Kerr coefficient over the range $-5~\mathrm{MHz}\lesssim K/(2\pi)\lesssim 6~\mathrm{MHz}$ and employed $K/(2\pi)=5.21~\mathrm{MHz}$ for the fast generation of Schr\"odinger cat states, establishing that Kerr nonlinearities on the several-megahertz scale are accessible in present circuit-QED devices~\cite{He2023}.
At the stronger-nonlinearity end, a high-impedance resonator operating in the effective single-junction limit, with the junction implemented as a flux-tunable SQUID, exhibited a $290~\mathrm{MHz}$ separation between the $\lvert 0\rangle\leftrightarrow\lvert 1\rangle$ and $\lvert 1\rangle\leftrightarrow\lvert 2\rangle$ transition frequencies. 
In the Kerr convention $K\hat{n}^{2}$ adopted here, this measured anharmonicity corresponds to an effective Kerr coefficient $|K|/(2\pi)\simeq145~\mathrm{MHz}$~\cite{Andersson2025}.
Each auxiliary qubit may be coupled to a dedicated lossy reset resonator, thereby providing an engineered relaxation channel~\cite{PhysRevLett.110.120501,Magnard2018PRL,Sunada2022PRApplied}. Alternatively, direct coupling to an open transmission line can provide a controllable radiative decay channel~\cite{Lu2021Decoherence,Brehm2021,Bolgar2020}.
When this engineered channel dominates the auxiliary-qubit energy relaxation, the experimentally measured population-decay rate corresponds directly to the coefficient $\gamma_\ell$ in the dissipator $\gamma_\ell\mathcal{D}[\hat{\sigma}_\ell^-]$. 
For example, a superconducting transmon directly coupled to an open transmission line exhibited a total energy-relaxation rate corresponding to $\gamma_\ell/(2\pi)\simeq0.28~\mathrm{MHz}$ in the notation used here~\cite{Lu2021Decoherence}, while more strongly overcoupled waveguide-QED devices have demonstrated radiative relaxation rates corresponding to $\gamma_\ell/(2\pi)\simeq6.4~\mathrm{MHz}$ on average and $\simeq8~\mathrm{MHz}$, respectively~\cite{Brehm2021,Bolgar2020}.
These measurements demonstrate that auxiliary-qubit relaxation rates ranging from approximately $0.3~\mathrm{MHz}$ to several megahertz are accessible in superconducting-circuit platforms.
The cavity-loss rate $\kappa$ can likewise be maintained in the kilohertz regime in superconducting nonlinear resonators. 
For example, a SNAIL-terminated resonator exhibited energy-relaxation times ranging from approximately $8$ to $20~\mu\mathrm{s}$ over the measured photon-number range, corresponding to $\kappa/(2\pi)\simeq0.008$--$0.020~\mathrm{MHz}$ in the notation used here~\cite{Lu2023SNAIL}. 
Together with a recent demonstration of a superconducting Kerr oscillator with a single-photon lifetime of $395~\mu\mathrm{s}$, corresponding to a cavity decay rate $\kappa/(2\pi)=0.40~\mathrm{kHz}$, these measurements indicate that the representative value $\kappa/(2\pi)=0.005~\mathrm{MHz}$ is experimentally accessible~\cite{Cai2026KerrSqueezing}.
Taken together, these experimentally demonstrated ingredients support the feasibility of implementing the joint cavity--qubit open-system model considered here, with engineered qubit relaxation supplying the irreversible reset required for the dissipation-rectified photon-raising cycle.

We next derive the parametrically activated AJC interaction used in Eq.\,\eqref{eq:sm_H}.
As illustrated in Fig.\,1(a) of the main text, each auxiliary qubit is coupled to the Kerr cavity through a tunable transverse coupling element whose coupling strength is modulated by an individual blue-sideband drive~\cite{Hoffman2011,Lu2017ParametricCoupling}. 
A minimal effective laboratory-frame Hamiltonian describing the resulting parametrically modulated cavity--qubit couplings is
\begin{align}
\hat H_{\rm lab}(t)
=
\omega_c\hat n
+K\hat n^2
+\sum_{\ell=0}^{N_q-1}
\omega_\ell
\hat\sigma_\ell^+\hat\sigma_\ell^-
+
\sum_{\ell=0}^{N_q-1}
2g_\ell\cos(\omega_{d,\ell}t+\phi_\ell)
(\hat a+\hat a^\dagger)
(\hat\sigma_\ell^++\hat\sigma_\ell^-),
\label{eq:sm_experimental_Hlab}
\end{align}
where $\omega_c$ is the linear cavity frequency, $\omega_\ell
$ is the transition frequency of qubit $\ell$, and $\omega_{d,\ell}$ and $\phi_\ell$ are the angular frequency and phase of the corresponding blue-sideband modulation drive. 
The first two terms describe the linear cavity mode and its self-Kerr nonlinearity, the third term gives the bare energies of the auxiliary qubits, and the last term represents the parametrically activated transverse cavity--qubit couplings. 
The prefactor $2g_\ell$ is chosen such that the resonant sideband interaction obtained after the rotating-wave approximation has coupling strength $g_\ell$.

To identify the resonant sideband interaction generated by the modulation, we move to the rotating frame defined by $\hat U(t)=\exp[-i t(\omega_c\hat n+\sum_{\ell=0}^{N_q-1}(\omega_{d,\ell}-\omega_c)\hat\sigma_\ell^+\hat\sigma_\ell^-)]$.
The Hamiltonian in this frame is
\begin{equation}
\hat H_{\rm rot}(t)
=
\hat U^\dagger(t)\hat H_{\rm lab}(t)\hat U(t)
-i\hat U^\dagger(t)\dot{\hat U}(t).
\label{eq:sm_experimental_Hrot_definition}
\end{equation}
Since the Kerr term commutes with $\hat n$, it remains unchanged under this transformation. The cavity and qubit operators transform according to
\begin{align}
\hat U^\dagger(t)\hat a\hat U(t)
&=
\hat a e^{-i\omega_c t},
\\
\hat U^\dagger(t)\hat\sigma_\ell^+\hat U(t)
&=
\hat\sigma_\ell^+
e^{i(\omega_{d,\ell}-\omega_c)t}.
\label{eq:sm_experimental_operator_transformations}
\end{align}
Using
\begin{equation}
2\cos(\omega_{d,\ell}t+\phi_\ell)
=
e^{i(\omega_{d,\ell}t+\phi_\ell)}
+
e^{-i(\omega_{d,\ell}t+\phi_\ell)},
\end{equation}
the transformed Hamiltonian can be written as
\begin{align}
\hat H_{\rm rot}(t)
=
K\hat n^2
+
\sum_{\ell=0}^{N_q-1}
\Delta_\ell
\hat\sigma_\ell^+\hat\sigma_\ell^-
+
\sum_{\ell=0}^{N_q-1}
g_\ell
\left(
e^{-i\phi_\ell}
\hat a^\dagger\hat\sigma_\ell^+
+
e^{i\phi_\ell}
\hat a\hat\sigma_\ell^-
\right)
+
\hat H_{\rm fast}(t),
\label{eq:sm_experimental_Hrot}
\end{align}
where
\begin{equation}
\Delta_\ell
=
\omega_\ell
+\omega_c-\omega_{d,\ell},
\label{eq:sm_experimental_detuning}
\end{equation}
and the remaining rapidly oscillating terms are
\begin{align}
\hat H_{\rm fast}(t)
=&
\sum_{\ell=0}^{N_q-1}g_\ell
\Big\{
e^{i(2\omega_{d,\ell}t+\phi_\ell)}
\hat a^\dagger\hat\sigma_\ell^+
+
e^{-i(2\omega_{d,\ell}t+\phi_\ell)}
\hat a\hat\sigma_\ell^-
\nonumber\\
&\quad+
e^{i(2\omega_ct+\phi_\ell)}
\hat a^\dagger\hat\sigma_\ell^-
+
e^{-i(2\omega_ct+\phi_\ell)}
\hat a\hat\sigma_\ell^+
\nonumber\\
&\quad+
e^{-i[2(\omega_{d,\ell}-\omega_c)t+\phi_\ell]}
\hat a^\dagger\hat\sigma_\ell^-
+
e^{i[2(\omega_{d,\ell}-\omega_c)t+\phi_\ell]}
\hat a\hat\sigma_\ell^+
\Big\}.
\label{eq:sm_experimental_Hfast}
\end{align}
For a modulation frequency close to the cavity--qubit sum frequency, $\omega_{d,\ell}\simeq\omega_c+\omega_\ell
$, the terms $\hat a^\dagger\hat\sigma_\ell^+$ and $\hat a\hat\sigma_\ell^-$ are slowly varying, whereas the terms in Eq.\,\eqref{eq:sm_experimental_Hfast} oscillate at frequencies of order $2\omega_{d,\ell}$, $2\omega_c$, or $2(\omega_{d,\ell}-\omega_c)$. 
Provided that
\begin{equation}
\max\left\{
g_\ell\sqrt{n+1},
|\Delta_\ell|,
|K|(2n+1),
\kappa,
\gamma_\ell
\right\}
\ll
\min\left\{
\omega_c,
|\omega_{d,\ell}-\omega_c|,
\omega_{d,\ell}
\right\}
\label{eq:sm_experimental_carrier_RWA}
\end{equation}
for the photon numbers relevant to the stabilization dynamics, these rapidly oscillating terms can be neglected within the rotating-wave approximation. 
The resulting effective Hamiltonian is
\begin{align}
\hat H_{\rm eff}
=
K\hat n^2
+
\sum_{\ell=0}^{N_q-1}
\Delta_\ell
\hat\sigma_\ell^+\hat\sigma_\ell^-
+
\sum_{\ell=0}^{N_q-1}
g_\ell
\left(
e^{-i\phi_\ell}
\hat a^\dagger\hat\sigma_\ell^+
+
e^{i\phi_\ell}
\hat a\hat\sigma_\ell^-
\right).
\label{eq:sm_experimental_Heff_phase}
\end{align}
Finally, defining
$\tilde{\sigma}_\ell^+
=e^{-i\phi_\ell}\hat\sigma_\ell^+$
and
$\tilde{\sigma}_\ell^-
=e^{i\phi_\ell}\hat\sigma_\ell^-$
absorbs the modulation phase into the qubit operators. Dropping the tildes then gives
\begin{equation}
\hat H_{\rm eff}
=
K\hat n^2
+
\sum_{\ell=0}^{N_q-1}
\Delta_\ell
\hat\sigma_\ell^+\hat\sigma_\ell^-
+
\sum_{\ell=0}^{N_q-1}
g_\ell
\left(
\hat a^\dagger\hat\sigma_\ell^+
+
\hat a\hat\sigma_\ell^-
\right),
\label{eq:sm_experimental_Heff}
\end{equation}
which reproduces Eq.\,\eqref{eq:sm_H}.

The separation of frequency scales required by Eq.\,\eqref{eq:sm_experimental_carrier_RWA} is compatible with experimentally demonstrated circuit-QED parameters.
Superconducting resonators and qubits typically operate at frequencies of several gigahertz, while experiments on parametrically activated blue-sideband interactions have employed modulation frequencies of order $\omega_{d,\ell}/(2\pi)\sim10~\mathrm{GHz}$~\cite{Roth2017}.
For qubit $\ell$, the modulation frequency is chosen to make the assigned AJC transition $|\ell,g_\ell\rangle\leftrightarrow|\ell+1,e_\ell\rangle$ resonant. Combining the rotating-frame detuning $\Delta_\ell=\omega_\ell
+\omega_c-\omega_{d,\ell}$ with the number-selective condition $\Delta_\ell=-(2\ell+1)K+\delta_{\rm com}$ gives
\begin{equation}
\omega_{d,\ell}
=
\omega_c+\omega_\ell
+(2\ell+1)K
-\delta_{\rm com}.
\label{eq:sm_experimental_drive_condition}
\end{equation}
For exact resonance, $\delta_{\rm com}=0$, and the representative values $\omega_c/(2\pi)=5~\mathrm{GHz}$, $\omega_\ell
/(2\pi)=6~\mathrm{GHz}$, and $K/(2\pi)=10~\mathrm{MHz}$ yield
\begin{equation}
\frac{\omega_{d,\ell}}{2\pi}
=
11~\mathrm{GHz}
+
(2\ell+1)\times10~\mathrm{MHz}.
\label{eq:sm_experimental_drive_frequencies}
\end{equation}
For four auxiliary qubits, this gives $\omega_{d,\ell}/(2\pi)=11.010$, $11.030$, $11.050$, and
$11.070~\mathrm{GHz}$ for $\ell=0,1,2,3$, respectively.
For the largest target considered, $N_q=4$, the reduced population model retains Fock states up to $n=N_{\mathrm{max}}=6$. 
Over this range, the largest Kerr-induced scale is $|K|(2n+1)/(2\pi)\leq130~\mathrm{MHz}$, while the detunings satisfy $|\Delta_\ell|/(2\pi)\leq70~\mathrm{MHz}$. 
These scales remain well below the gigahertz-scale cavity, qubit, and modulation frequencies. For $g_\ell/(2\pi)=0.1~\mathrm{MHz}$, the corresponding coupling matrix elements satisfy $g_\ell\sqrt{n+1}/(2\pi)\lesssim0.27~\mathrm{MHz}$, while the dissipative rates remain in the kilohertz-to-megahertz range. 
The rotating-wave approximation used to obtain Eq.\,\eqref{eq:sm_H} is therefore well justified over the truncated Fock space considered here.

\end{document}